\documentclass[prl,twocolumn,superscriptaddress,showpacs, letterpaper]{revtex4-1}

\usepackage{graphicx} 
\usepackage{dcolumn}   
\usepackage{float}
\usepackage{wrapfig}
\usepackage{verbatim}
\usepackage{amsmath}
\usepackage{bm}        
\usepackage{amssymb}
\usepackage{amsfonts}
\usepackage{textcomp}
\usepackage{accents}
\usepackage{fancyhdr}
\usepackage{euscript}
\usepackage{blkarray}
\usepackage{multirow}
\usepackage{tikz}
\usepackage{subfigure}
\usepackage[nolist,nohyperlinks]{acronym}
\usepackage{enumitem}
\newcommand{\bea}{\begin{eqnarray}}
\newcommand{\eea}{\end{eqnarray}}
\newcommand{\be}{\begin{equation}}

\newcommand{\ee}{\end{equation}}
\newcommand{\ba}{\begin{align}}
\newcommand{\ea}{\end{align}}
\newcommand{\bi}{\begin{itemize}}
\newcommand{\ei}{\end{itemize}}
\newcommand{\bt}{\begin{tabular}}
\newcommand{\e}{\end{tabular}}
\newcommand{\nn}{\nonumber}

\newcommand{\vl}{\kern-\arraycolsep & \kern-\arraycolsep}
\newcommand{\VL}{\kern-\arraycolsep\strut\vrule &\kern-\arraycolsep}
 
\newcommand{\G}{\Gamma}
\newcommand{\x}{\xi}
\newcommand{\ztwos}{(\nu_0;\nu_1\nu_2\nu_3)}
\newcommand{\ztwo}{{\mathcal Z}_2}

\begin{document}

\title{Dirac Line Nodes in Inversion Symmetric Crystals }

\author{Youngkuk Kim}
\affiliation{The Makineni Theoretical Laboratories, Department of Chemistry, University of Pennsylvania, Philadelphia, Pennsylvania 19104--6323, USA}
\author{Benjamin J. Wieder}
\affiliation{Department of Physics and Astronomy, University of Pennsylvania, Philadelphia, Pennsylvania 19104--6396, USA}
\author{C. L. Kane}
\affiliation{Department of Physics and Astronomy, University of Pennsylvania, Philadelphia, Pennsylvania 19104--6396, USA}
\author{Andrew M. Rappe}
\affiliation{The Makineni Theoretical Laboratories, Department of Chemistry, University of Pennsylvania, Philadelphia, Pennsylvania 19104--6323, USA}
\date{\today}

\begin{abstract}
We propose and characterize a new $\mathbb{Z}_2$ class of topological semimetals with a vanishing spin--orbit interaction.  The proposed topological semimetals are characterized by the presence of bulk one-dimensional (1D) Dirac Line Nodes (DLNs) and two-dimensional (2D) nearly-flat surface states, protected by inversion and time--reversal symmetries. We develop the $\mathbb{Z}_2$ invariants dictating the presence of DLNs based on parity eigenvalues at the parity--invariant points in reciprocal space. Moreover, using first-principles calculations, we predict DLNs to occur in Cu$_3$N near the Fermi energy by doping non-magnetic transition metal atoms, such as Zn and Pd, with the 2D surface states emerging in the projected interior of the DLNs. This paper includes a brief discussion of	 the effects of spin--orbit interactions and symmetry-breaking as well as comments on experimental implications.
\end{abstract}

\maketitle
A recent development in condensed matter physics has been the discovery of semimetallic features in electronic band structures protected by the interplay of symmetry and topology.    A tremendous amount of progress has been made in materials with strong spin--orbit interactions, such as the surface states of topological insulators  \cite{Hasan10p3045, Qi11p1057} and topological crystalline insulators \cite{Fu11p106802}, as well as the gapless bulk states of Weyl and Dirac semimetals \cite{Wan11p205101, Steve12p140405, Steinberg14p036403}.      Related topological phenomena can occur in materials with vanishing (or weak) spin--orbit interactions \cite{Alexandradinata14p116403}.   Indeed, the prototypical topological semimetal is graphene \cite{Neto09p109}, which exhibits Dirac points that are robust to the extent that the spin--orbit interaction in carbon is weak.   In the absence of spin--orbit interactions, the Dirac points in graphene are topologically protected by the combination of inversion and time--reversal symmetries.   

In this paper we study a related phenomenon for three dimensional (3D) materials with weak spin--orbit interaction.   We show that the combination of inversion and time--reversal symmetries protects Dirac line nodes (DLNs), for which the conduction band and valence band meet along a line in momentum space, and we predict realistic materials in which they should occur.  DLNs have been discussed previously in the context of models that have an additional chiral symmetry, which can arise on a bipartite lattice with only nearest neighbor hopping.   In this case, the DLN can be constrained to occur at zero energy.  However, chiral symmetry is never expected to be an exact symmetry of a band structure.   We will show that despite the absence of chiral symmetry, the line node is protected, though it is not constrained to sit at a constant energy.  We will show, however, that in the vicinity of a band inversion transition, a DLN can occur in the form of a small circle, whose energy is approximately flat.   The presence of such a Dirac circle has interesting consequences for the surface states, and we show that on the projected interior of the Dirac circle, the surface exhibits a nearly flat band, which must be half--filled when the surface is electrically neutral.    Such surface states could be an interesting platform for strong correlation physics.  We introduce a class of materials and use first--principles density functional theory (DFT) calculations to show that they can be tuned through the band inversion transition and exhibit the predicted Dirac circle, as well as a more complex nodal structure.  A similar DLN near an inversion transition has recently been predicted in a 3D graphene network \cite{Weng14p1, Volovik11p1}.  Recently, DLNs also have been proposed in systems with strong spin-orbit interactions, such as perovskite irridates \cite{Kim14p1, Chen15p1} and non-centrosymetric semimetals \cite{Weng15p011029}, but in those systems the mechanism of symmetry protection is different.

We will begin by elucidating the topological constraints that inversion and time--reversal symmetries impose.  We will then introduce a $\mathbb{Z}_2$ topological invariant (related to the invariant characterizing a 3D topological insulator) which signifies the presence of DLNs.     We will then present DFT calculations on transition metal--doped Cu$_3$N that predict a Dirac circle, as well as nearly-flat boundary modes.    We will then introduce a simple low--energy $k\cdot p$ model that explains the appearance of the Dirac circle at a band inversion, and allows for a simple description of the resulting boundary modes.   

Consider a 3D Bloch Hamiltonian $\mathcal{H}({\bf k})$ that is invariant under inversion $\mathcal{P}$ and time--reversal $\mathcal{T}$.   In the absence of spin--orbit interactions we may consider $\mathcal{T}^2=+1$.   The occupied Bloch eigenstates are characterized  by a Berry connection ${\bf A}({\bf k}) = -i \sum_n \langle u_n({\bf k})|\nabla_{\bf k} u_n({\bf k})\rangle$.   $\mathcal{P}$  and $\mathcal{T}$ symmetries constrain the Berry phase, $\omega(C) = \exp i \oint_C {\bf A}\cdot d{\bf k}$, on any closed loop $C$ in momentum space, to satisfy $\omega(C) = \omega(-C)$ and $\omega(C)= \omega(-C)^*$, respectively.  It  follows that loops $C$ are characterized by a $\mathbb{Z}_2$ topological invariant $\omega(C) = \pm 1$ \footnote{Since $[\mathcal{H}({\bf k}),\mathcal{P}\mathcal{T}]=0$, this $\mathcal{Z}_2$ invariant can also be understood as characterizing one parameter families of Hamiltonians in class AI \cite{Teo10p115120}}. The non--trivial loops $\omega(C) = -1$ must enclose a degeneracy.   In two dimensions, this explains the symmetry protection of the Dirac points in graphene.   In three dimensions, it guarantees that a line of degeneracies must pierce any surface bounded by $C$.   

The parity eigenvalues $\xi_n(\Gamma_a) = \pm 1$ of the occupied Bloch states at the 8 parity--invariant momenta $\Gamma_a$ provide an important constraint that allows us to identify topologically protected line nodes.  First, consider a time-reversal invariant loop $C_{ab}= c_{ab} - \bar c_{ab}$ that connects $\Gamma_a$ and $\Gamma_b$ along two time--reversed paths $c_{ab}$ and $\bar c_{ab}$.  In the supplementary material we prove that the Berry phase on this loop satisfies
\begin{equation}
\omega(C_{ab}) = \xi_a \xi_b;  \quad\quad \xi_a = \prod_n \xi_n(\Gamma_a).
\end{equation}
If we now consider four parity-invariant points, the contour $C_{ab}-C_{cd}$ defines the boundary $\partial S_{abcd}$ of a surface $S_{abcd}$ in momentum space.  The Berry phase on $\partial S_{abcd}$ counts the number of DLNs $N(S_{abcd})$ that pierce that surface.   We thus conclude that
\begin{equation}
(-1)^{N(S_{abcd}) }= \xi_a \xi_b \xi_c \xi_d.
\end{equation}
Thus, when $\xi_a \xi_b \xi_c \xi_d=-1$ there must be an odd number of DLN piercing any surface $S_{abcd}$, with the simplest case being just a single one. This relation is quite similar to the topological invariants $(\nu_0;\nu_1\nu_2\nu_3)$ characterizing a (strong or weak) $\mathbb{Z}_2$ topological insulator in the presence of spin--orbit interactions \cite{Fu07p106803, Fu07p045302}. Indeed, in a topological insulator with $\xi_a \xi_b \xi_c \xi_d = -1$ , when the spin--orbit interaction is turned off a DLN must appear, because the system can not be adiabatically connected to a trivial insulator. 

This connection to the parity eigenvalues suggests a route towards realizing the DLNs:   Starting with a trivial insulator, invert a pair of opposite-parity bands.     At the inversion transition a small Dirac circle will necessarily emerge and grow.  In the following we will predict a class of real materials which exhibits this behavior, and analyze the low--energy structures which emerge.

\begin{figure}[tb!]
\includegraphics[width=0.30\textwidth]{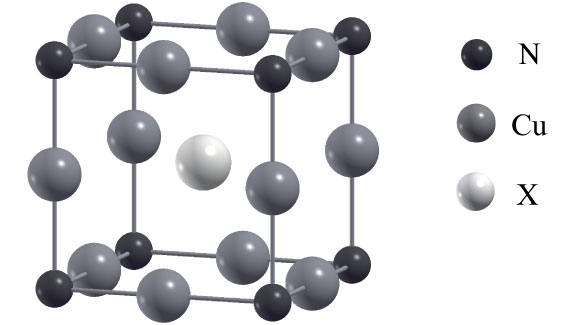}
\caption{\label{fig1}
Crystal structure of Cu$_3$NX. X represents a transition metal atom intercalated at the body--center of the cubic unit cell of Cu$_3$N in an anti--ReO$_3$ structure. }
\end{figure}

Searching for materials that consist of light elements and preserve $\mathcal{T}$ and $\mathcal{P}$ symmetries, we find that  copper nitride, Cu$_3$N, a narrow--gap semiconductor ($E_g \sim 0.3$ eV) \cite{Juza38p282}, fosters DLNs near the Fermi level via an insulator--to--metal transition driven by doping transition metal atoms. Copper nitride, first synthesized in 1937 \cite{Juza38p282}, is stable in air at room temperature with a cubic anti--ReO$_3$ structure in space group 221 (Pm$\bar{3}$m).   It contains a rather large void at the center of the cubic unit cell, as shown in Fig.\ \ref{fig1}. This void can host extrinsic atoms such as N \cite{Hadian12p1067}, Li \cite{Gulo04p2032, Hou08p1651}, Pd \cite{MorenoArmenta07p166, Zachwieja90p175, Hahn96p12684, Ji06p252120, Sieberer06p014416}, Rh, Ru \cite{Sieberer06p014416}, Zn, Ni, Cd \cite{MorenoArmenta07p166}, Cu \cite{Hou08p1651, MorenoArmenta04p9}, Fe, Ti \cite{Fan07p254}, Ag \cite{Pierson08p568}, La, Ce \cite{Wu14p221}, as well as many other transition--metal atoms \cite{Cui12p3138}. In particular Ni, Cu, Pd, An, Ag, and Cd \cite{MorenoArmenta07p166} are found to drive an electronic transition in Cu$_3$N into a semimetal without breaking $\mathcal{T}$ symmetry \cite{Cui12p3138}, by which we expect that DLNs form near the Fermi energy.

To demonstrate the existence of DLNs in the transition metal--doped Cu$_3$N, we perform first--principles calculations based on DFT. We employ the Perdew--Burke--Ernzerhof--type generalized gradient approximation \cite{Perdew96p3865} as implemented in the QUANTUM ESPRESSO package \cite{Giannozzi09p395502}.  Norm--conserving, optimized, designed nonlocal pseudopotentials are generated by the OPIUM package \cite{Rappe90p1227, Ramer99p12471}.  The wave functions are expanded in a plane--wave basis with an energy cutoff of 680 eV.  We initially consider the spin--orbit interaction based on a scalar--relativistic pseudopotential \cite{Ilya00p2311}, and later, we will discuss the effect of spin--orbit interactions, based on a fully-relativistic non--collinear scheme.

\begin{figure}[tb]
\includegraphics[width=0.48\textwidth]{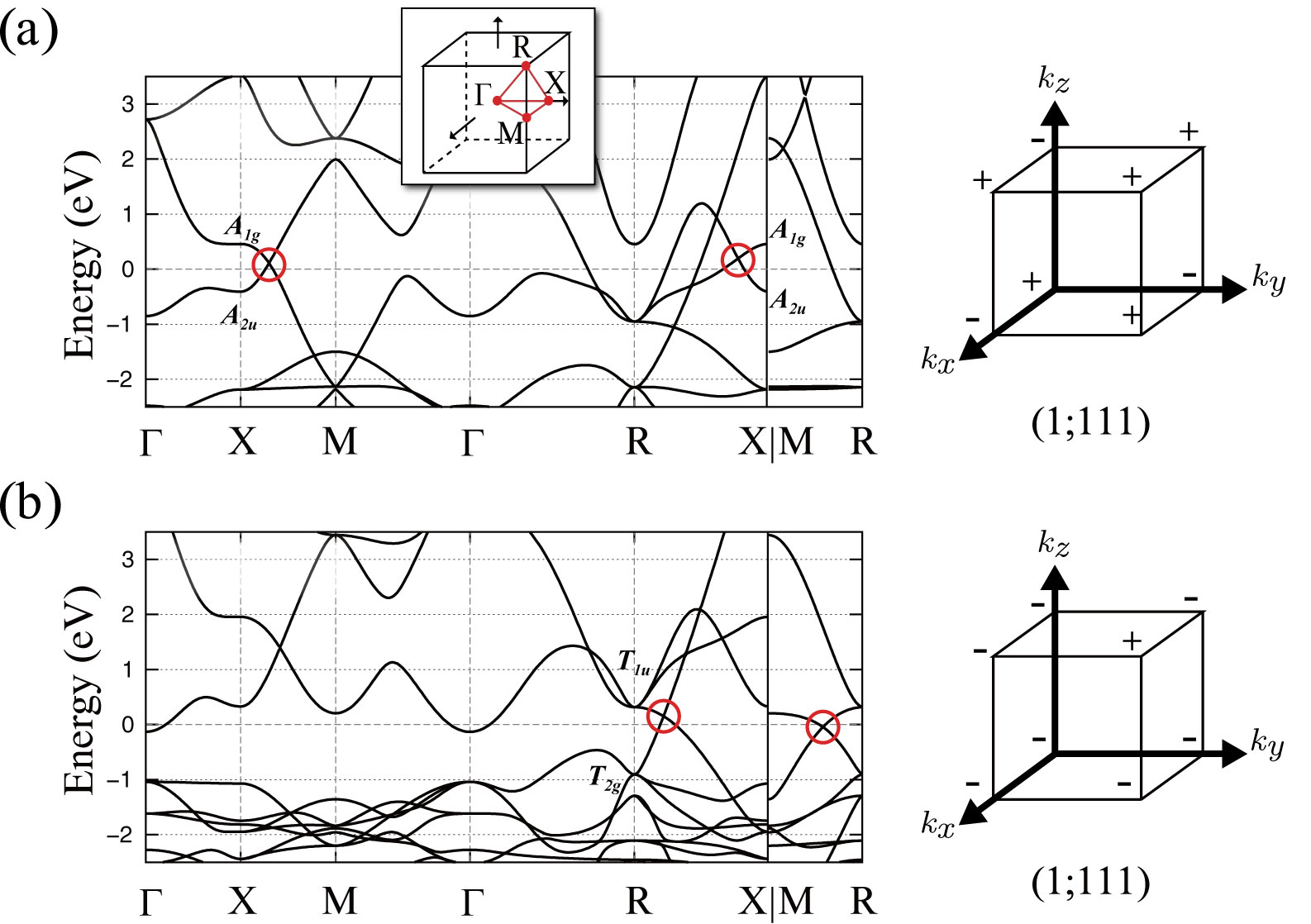}
\caption{\label{fig2} (color online)
Electronic structures and $\mathbb{Z}_2$ indices of (a) Cu$_3$NZn and (b) Cu$_3$NPd. Bands are drawn along the high--symmetry lines of the BZ (inset). The Dirac points are indicated by red circles. Parity eigenvalues are illustrated at the eight parity--invariant points in the first octant of the BZ.} 	
\end{figure}

The low--energy electronic structures of Cu$_3$NX are more or less similar for X $=$ \{Ni, Cu, Pd, An, Ag, Cd\}, as reported in Ref. \cite{MorenoArmenta07p166}.   Here we present the results of Cu$_3$NZn and Cu$_3$NPd as representatives of transition metal--doped Cu$_3$N systems. Note that these are extreme cases where the transition metal atoms are maximally doped \footnote{See the supplementary material for the discussion on the doping concentration.}. In Cu$_3$NZn the conduction and valence bands are mainly comprised of conduction $A_{2u}$ and valence $A_{1g}$ states near the Fermi energy. As shown in Fig.\ \ref{fig2}, these bands are inverted at the $X$ points, forming two--dimensional (2D) Dirac points on the $X$--$M$ and $R$--$X$ lines (enclosed by red circles in the figure). These Dirac points signal the presence of a DLN enclosing $X$.  Although there are more degenerate points near the Fermi level, and bands crossing the Fermi energy near the R point, we will simplify and here focus only on the bands near $X$. On the other hand, the conduction and valence bands of Cu$_3$NPd are comprised of $T_{2g}$ and $T_{1u}$ states, which are inverted at the $R$ point, forming the Dirac points on the $R$--$X$ and $M$--$R$ lines.  These Dirac points are in fact parts of a DLN that encloses the $R$ point, as shown below.

\begin{figure}[tb]
\includegraphics[width=0.48\textwidth]{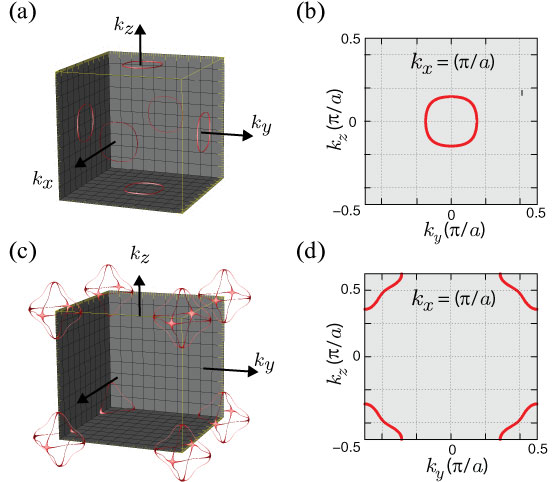}
\caption{\label{fig3} (color online) Dirac line nodes in the Brillouin Zone (BZ).  (a) and (b) Cu$_3$NZn, and (c) and (d) Cu$_3$NPd. The DLNs are illustrated by red curves in the 3D BZ [(a) and (c)] and on the 2D boundary plane of the BZ at $k = X^x$ [(b) and (d)].}
\end{figure}

The nodal lines of the conduction and valence bands in the 3D BZ are shown in Fig.\ \ref{fig3}.  As mentioned above, DLNs appear near the $X$ points in the Cu$_3$NZn system.  The cubic symmetry of the system dictates three DLNs encircling the three inequivalent $X$ points $X^r = \pi \hat{r}/a$, where $r = x, y$, and $z$.  Similarly, Cu$_3$NPd also exhibits three DLNs due to the cubic symmetry, but since they appear enclosing the $R$ point, they form in a gyroscope shape.  In both systems, the DLNs are contained in three mirror-invariant planes at $X^x$, $X^y$, and $X^z$, due to the corresponding mirror symmetries.  We expect that breaking the mirror symmetries should unlock the DLNs from the mirror planes, but that the DLNs will survive as they are protected by $\mathcal{P}$ and $\mathcal{T}$.

The appearance of DLNs agrees with the topological prediction of $\mathbb{Z}_2$ invariants $(\nu_0;\nu_1\nu_2\nu_3)$, calculated from the parity analysis.  In Cu$_3$NZn, parities at the eight time--reversal invariant momenta ($\Gamma,3X,3M,R$) give $(\nu_0;\nu_1\nu_2\nu_3) = (1;111)$, which dictates that there should be DLNs threading half the invariant plane at $X^r = \pi \hat{r}/a$ ($r = x, y, z$) an odd number of times.  The three DLNs enclosing the $X$ points fulfill this topological constraint (see the supplementary material for more details of this analysis).  Similarly, in Cu$_3$NPd we find that $(\nu_0;\nu_1\nu_2\nu_3) = (1;111)$, which is also in accordance with the formation of the three DLNs enclosing $R$. In this case, each invariant plane at $X^r$ is threaded three times by all three DLNs. 

A low--energy $\boldsymbol{k}\cdot\boldsymbol{p}$ Hamiltonian describing the conduction $A_{2u}$ and valence $A_{1g}$ states, which form the DLNs in Cu$_3$NZn, captures the essential features of the DLNs. Near $X^r$, symmetries dictate a two--band Hamiltonian
\begin{equation}
\label{eqn:H}
   \mathcal{H}_r = ( \bar{\epsilon} + a_{\perp} |\boldsymbol{q}_{\perp}|^2 + a_{r}  q_{r}^2 ) \mathbb{I}_\tau     + v q_r \tau^y 
   + ( \Delta \epsilon + b_\perp
   |\boldsymbol{q}_\perp|^2 + b_r q_{r} ^2)\tau^z,
\end{equation}
where $\boldsymbol{q} = \boldsymbol{k} - X^r$, $\perp$ represents the normal components to $\hat{r}$,  and the Pauli matrices  $\{ \mathbb{I}_\tau,\tau^i \}$ describe the $A_{1g}$ and $A_{2u}$ states.  The form of $\mathcal{H}_r$ is uniquely determined by inversion $\mathcal{P} =  \tau^z$ and time--reversal $\mathcal{T} = K$ ($K$ being complex conjugation), together with the $D_{4h}$ point group symmetries of $X$.  It gives energy eigenvalues
\begin{eqnarray}
   E_\pm (\boldsymbol{q}) = \bar{\epsilon} &+&  a_{\perp} |\boldsymbol{q}_{\perp}|^2 + a_{r}  q_{r}^2   \nn \\
    &\pm &   \sqrt{( \Delta \epsilon + b_\perp |\boldsymbol{q}_\perp|^2 + b_r q_{r} ^2)^2  + v^2q_r^2}.
\end{eqnarray}
A DLN forms at $q_r = 0 $ and $|\boldsymbol{q}_\perp| ^2 \equiv q_0^2 =	-\Delta/b_\perp$, when the bands are inverted ($\Delta \epsilon< 0$). The DFT results determine $\Delta \epsilon 	\sim - 0.4$ eV.  In Cu$_3$NPd, unlike in Cu$_3$NZn, there are conduction $T_{2g}$ and valence $T_{1u}$ states, instead leading to a six--band Hamiltonian $\mathcal{H}$.  However, this can be decomposed into three copies of $\mathcal{H}_r$ with $r = x, y$, and $z$ and $\mathcal{H} = \mathcal{H}_x \oplus \mathcal{H}_y \oplus \mathcal{H}_z$, giving rise to three gyroscope--shaped DLNs.  Therefore, the essential features of the DLNs should be the same between Cu$_3$NZn and Cu$_3$NPd, aside from the former having a single DLN occurring in three inequivalent valleys of the BZ ($X$ points) and the latter having three DLNs in a single valley ($R$ point).

This model Hamiltonian also describes boundary modes. Consider a boundary perpendicular to $\hat{r}$ in which $\Delta \epsilon$ varies between a negative (inverted) value and a large positive value. Fixing $q_\perp$ and considering the theory to linear order in $q_r \rightarrow -i\partial_r$,
\begin{equation}
   \mathcal{H}_z(\boldsymbol{q}) = 
   -iv \tau^y \partial_r
   + ( \Delta \epsilon(r) + b_\perp  q_\perp^2)\tau^z 
   + ( \bar{\epsilon} + a_{\perp} q_{\perp}^2 )\mathbb{I}_\tau.
\end{equation}
For each $k_\perp$ this defines a Jackiw--Rebbi problem \cite{Jackiw76p3398}. When $\Delta\epsilon + b_\perp k_\perp^2 < 0$ there will be a boundary mode at the surface. In general the boundary band is not flat, but disperses for $k_\perp < k_\mathrm{F}$ 
\begin{equation}
\epsilon_0 (k_\perp) = \bar{\epsilon} + a_\perp k^2_\perp \le 0.
\end{equation}
If $a = 0$ however, the surface band is flat. This reflects an additional chiral symmetry $\{\mathcal{H},\tau^x\} = 0$ at this point.  In this model, the value of $a$ is related to the difference of the effective masses of the $A_{1g}$ and $A_{2u}$ bands. If the surface in the absence of inversion is electrically neutral, then after inversion the surface will be neutral when the surface band is {\it half--filled}. This leads to a narrow surface band, where electron density  $q_0^2/4\pi  = |\Delta \epsilon|/4\pi b_\perp$ is controlled by the degree of band inversion. In the absence of screening from other bands, this surface band will tend to be pinned at the Fermi energy.

\begin{figure}[tb] 
\includegraphics[width=0.48\textwidth]{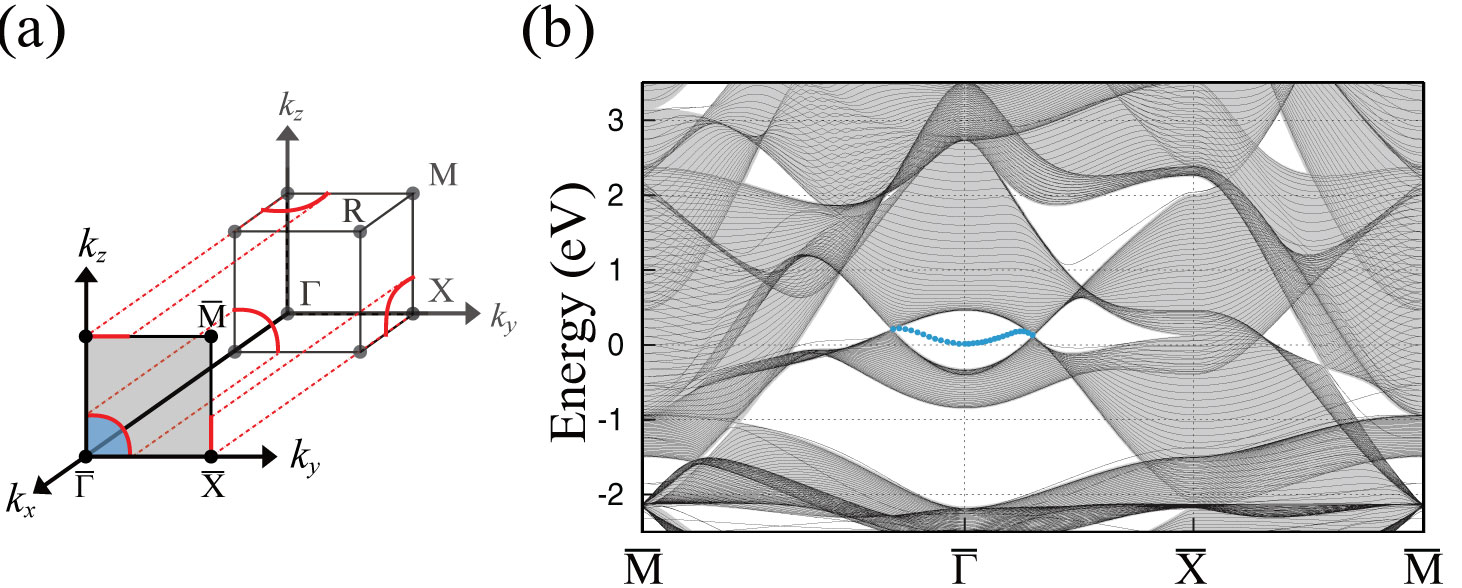}
\caption{\label{fig4} (color online)  Two--dimensional surface electronic structure for Cu$_3$NZn. (a) First octant of three--dimensional Brillouin zone (BZ) of Cu$_3$NZn projected onto the two--dimensional surface BZ of the (100) surface and (b) surface electronic band structure. The Dirac line nodes (DLNs) and the projected interior of DLNs are illustrated with the red and blue schemes, respectively. The slab bands are shown in black lines and surface states in the enclosed region are shown in blue lines. The shaded region represents bulk bands projected onto the BZ of the (100) surface along $k_x$	.}
\end{figure}    
To study the surface states in Cu$_3$NZn, we calculate the band structures of a slab geometry with 40 unit cells, exposing the (100) Cu$_2$N surfaces to vacuum. Our calculation from first principles predicts	that nearly-flat surface states emerge in the interiors of projected DLNs connecting the Dirac nodes, as shown in Fig.\ \ref{fig4}. The slab band structure exhibits the weakly-dispersing surface states near $\bar{\Gamma}$ in the projected interior of the DLN. The topological surface states resulting from closed DLNs are half--filled and nearly flat, providing a unique venue for interesting strong-correlation and transport physics.

The strong spin--orbit interaction can induce diverse topological phases in DLN semimetals, including topological insulators, 3D Dirac semimetals \cite{Steve12p140405, Wang13p125427}, or even other DLN semimetals \cite{Chen15p1}. Analogously to graphene, spin--orbit interaction can gap out DLNs and drive the system to a topologically-insulating phase. The resultant topological insulator should have the same topological $\mathbb{Z}_2$ indexes as the DLN semimetal from which it originated. More interestingly, an additional crystalline symmetry may protect a part of the DLN in a symmetry-invariant region of the BZ, resulting in topological Dirac semimetals or crystalline symmetry--protected DLNs with strong spin--orbit interactions. We have tested the effect of spin--orbit interaction in Cu$_3$NPd using a fully-relativistic non--collinear scheme, and indeed found that $C_4$ symmetry along the $R$--$M$ line protects the Dirac point on the line, while the spin--orbit coupling otherwise opens a gap (with maximum size of $\sim$ 62 meV on the $R$--$X$ line), thus giving rise to a 3D Dirac semimetal phase in a strong spin--orbit interacting regime. Note that Cu$_3$NPd is an extreme case where Pd is maximally doped, and thus the spin--orbit interactions due to Pd $4d$ states are maximized. The spin--orbit interaction can be controlled either by the Pd--doping concentration, or by doping other group-X transition-metal atoms, such as Ni, Pd, and Pt. We thus expect both the DLN semimetal and 3D Dirac semimetal phases should be accessible in the Cu$_3$N system.

In summary, we have demonstrated that the combination of inversion and time--reversal symmetries allows for the $\mathbb{Z}_2$ classification of topological semimetals under vanishing spin--orbit interactions. The proposed topological semimetals are characterized by the presence of bulk DLNs and nearly-flat surface states, protected by inversion and time-reversal symmetries. Our first--principles calculations predict that the proposed topological phase can be observed in Cu$_3$N by doping with a class of non-magnetic transition metal atoms X, where X $=$ \{Ni, Cu, Pd, Ag, Cd\}. The 2D surface states predicted for the DLN semimetal can hopefully be experimentally observed through, for example, ARPES in Cu$_3$NX$_x$, using the doping concentration $x$ as a knob to control the sizes of the closed DLN and the enclosed surface band. Doping with heavier atoms can also be used to potentially observe spin--orbit-induced topological phases.

\begin{acknowledgments}
While this manuscript was in the final stages of preparation we learned of recent work proposing DLN in Ca$_3$P$_2$ \cite{Xie15p1}. YK acknowledges support from NSF grant DMR--1120901. CLK acknowledges support from a Simons Investigator grant from the Simons Foundation. AMR acknowledge support from the DOE Office of Basic Energy Sciences, under grant number DE--FG02--07ER15920.  Computational support is provided by the HPCMO of the U.S. DOD and the NERSC of the U.S. DOE.
\end{acknowledgments}

\renewcommand{\thefigure}{S\arabic{figure}}
\setcounter{figure}{0}
\renewcommand{\theequation}{S\arabic{equation}}
\setcounter{equation}{0}

\section{Supplementary Material for Dirac Line Nodes in Inversion Symmetric Crystals}
\subsection{Proof of Eqn. (1).}
\begin{figure*}[tb!]
\includegraphics[width=0.80\textwidth]{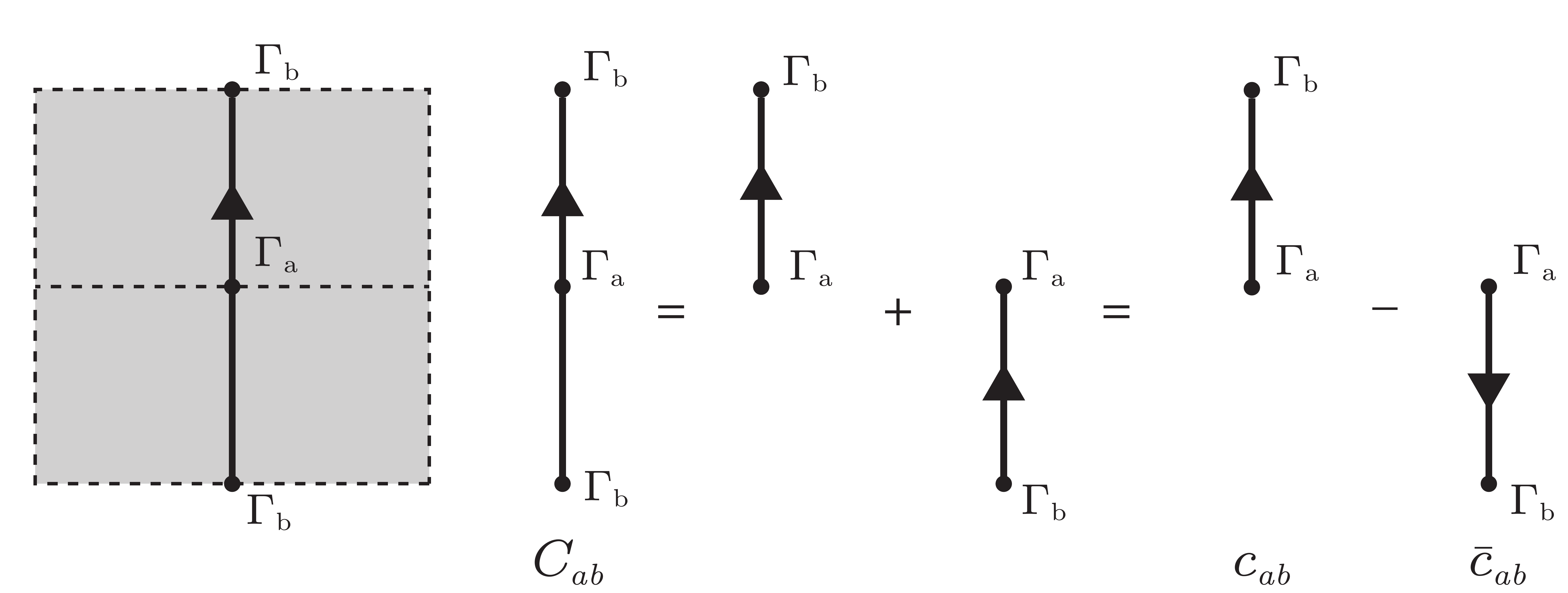}
\caption{\label{sfig1}
Invariant loop $C_{ab}$ in the Brillouin zone (BZ). The invariant loop $C_{ab}$  = $ c_{ab} - \bar c_{ab}$ connects two invariant points $\G_a$ and $\G_b$, where  $\bar c_{ab}$ is a path from $\G_a$ to $\G_b$ along the time--reverse of $c_{ab}$. }
\end{figure*}
Here we prove that the path integral of a  Berry connection ${\bf A}({\bf k}) = -i \sum_n \langle u_n({\bf k})|\nabla_{\bf k} u_n({\bf k})\rangle$ 
\begin{equation}
\omega(C_{ab}) = e^{ i \oint_{C_{ab}} {\bf A}\cdot d{\bf k}}
\end{equation}
on a loop $C_{ab} = c_{ab} - \bar{c}_{ab}$ that connects two parity-- and time--reversal--invariant points $\G_a$ and $\G_b$ along the two time--reversal paths $c_{ab}$ and $\bar c_{ab}$  (See Fig.\ \ref{sfig1}) can be obtained by the parity eigenvalues $\xi_n(\Gamma_a) = \pm 1$ of the occupied Bloch states at parity--invariant momenta $\Gamma_a$
\begin{equation}
\label{seqn2}
\omega(C_{ab}) = \xi_a \xi_b;  \quad\quad \xi_a = \prod_n \xi_n(\Gamma_a).
\end{equation}
It follows that
\begin{eqnarray}
  \omega (C_{ab}) &=& e^{i\left( \int_a^b {\bf A({\bf k })}\cdot d{\bf k}|_{c_{ab}} - \int_a^b {\bf A({\bf k })}\cdot d{\bf k}|_{\bar c_{ab}}\right)} \nn \\
                            &=& e^{i\int_a^b \left({\bf A}({\bf k}) - {\bf A}(-{\bf k})\right)\cdot d{\bf k}|_{c_{ab}}}. 
\end{eqnarray}
Inversion symmetry ($\mathcal{P}$) guarantees that 
$|u_n(-{\bf k})\rangle = e^{i\beta_n({\bf k})} \mathcal{P} |u_n({\bf k})\rangle.$
It then follows that 
${\bf A}({\bf k}) - {\bf A}(-{\bf k}) = \nabla_k \sum_n\beta_n({\bf k}),$
so that
\begin{equation}
\omega(C_{ab}) = e^{ i\sum_n\beta_n(\Gamma_b) - \beta_n(\Gamma_a)}.
\end{equation}
Now consider $P({\bf k}) = \prod_n \langle u_n(-{\bf k})|\mathcal{P} |u_n({\bf k})\rangle = e^{-i \sum_n\beta_n({\bf k})}$.
For a parity--invariant point $\Gamma_i$ , $P(\Gamma_i) = \prod_n \xi_n(\Gamma_i)$, where $\xi_n(\Gamma_i) = \langle u_n(\Gamma_i)|\mathcal{P}|u_n(\Gamma_i)\rangle$ is the parity eigenvalue.   We thus prove Eqn. (\ref{seqn2}).

\subsection{$\ztwo$ topological invariants and Dirac line nodes of Cu$_3$NZn}

\begin{figure*}[tb!]
\includegraphics[width=0.80\textwidth]{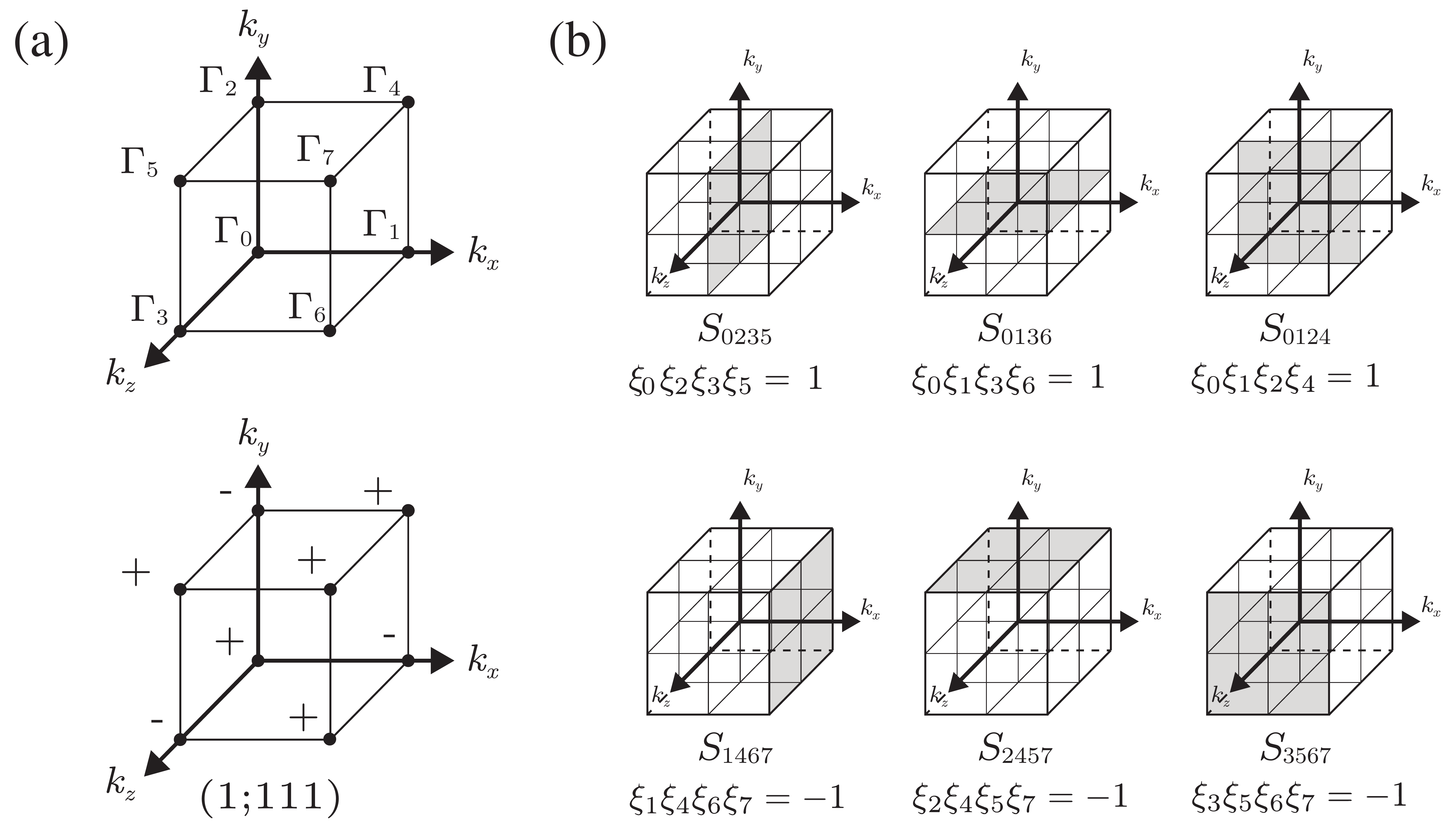}
\caption{\label{sfig2}
Six invariant surfaces $S_{abcd}$ in Cu$_3$NZn. (a) First octant of three--dimensional (3D) BZ. $\xi_i$ of Cu$_3$NZn are presented at parity--invariant momenta $\G_i$, which determine the $\ztwos = (1;111)$ phase. (b) Shaded regions represent three trivial invariant planes (top panel) and three nontrivial invariant planes (bottom panel).}
\end{figure*}
The topological invariant $\xi_a\xi_b\xi_c\xi_d = -1$ dictates that  for any invariant surface $S_{abcd}$ of the BZ, hosting four invariant momenta $\Gamma_i$ ($i=a,b,c,d$), there will be an odd number of Dirac line nodes (DLNs) intersecting the half surface at ${\bf k}$ (and the other half at $-{\bf k}$). Here we show that the DLNs that appear in Cu$_3$NZn satisfy this topological constraint. The cubic BZ of Cu$_3$NZn has eight distinct invariant momenta ($\Gamma,3X,3M,R$), and the parity eigenvalues at the high--symmetry momenta are calculated as $(\x(\Gamma), \x(X),\x(M),\x(R)) = (1,-1,1,1)$ as shown in Fig \ref{sfig2}.  Similar to the $\ztwo$ topological invariants of topological insulators \cite{Fu07p106803}, $\ztwo$ topological invariants $(\nu_0;\nu_1\nu_2\nu_3)$ can be defined in the DLN semimetals as 
\begin{equation}
(-1)^{\nu_0} = \prod_{n_j = 0,1} \xi_{n_1n_2n_3},
\end{equation} 
\begin{equation}
(-1)^{\nu_{i=1,2,3}} = \prod_{n_{j \ne i} = 0,1} \xi_{n_1n_2n_3},
\end{equation} 
where  $\xi_{i=(n_1n_2n_3)}$ are parity eigenvalues at the eight invariant momenta, $\Gamma_{i=(n_1n_2n_3)} = (n_1 {\bf b}_1 +n_2 {\bf b}_2 + n_3 {\bf b}_3)/2$, with $n_j = 0, 1$, and the primitive reciprocal lattice vectors ${\bf b}_i$.  The $\ztwo$ invariants in Cu$_3$NZn  are then obtained as $(\nu_0;\nu_1\nu_2\nu_3) = (1;111)$, which dictate that an odd number of DLNs will pierce half the nontrivial invariant surfaces, $S_{1467}$, $S_{2457}$, and $S_{3567}$ (See Fig.\ \ref{sfig2}).

\begin{figure*}[tb!]
\includegraphics[width=0.85\textwidth]{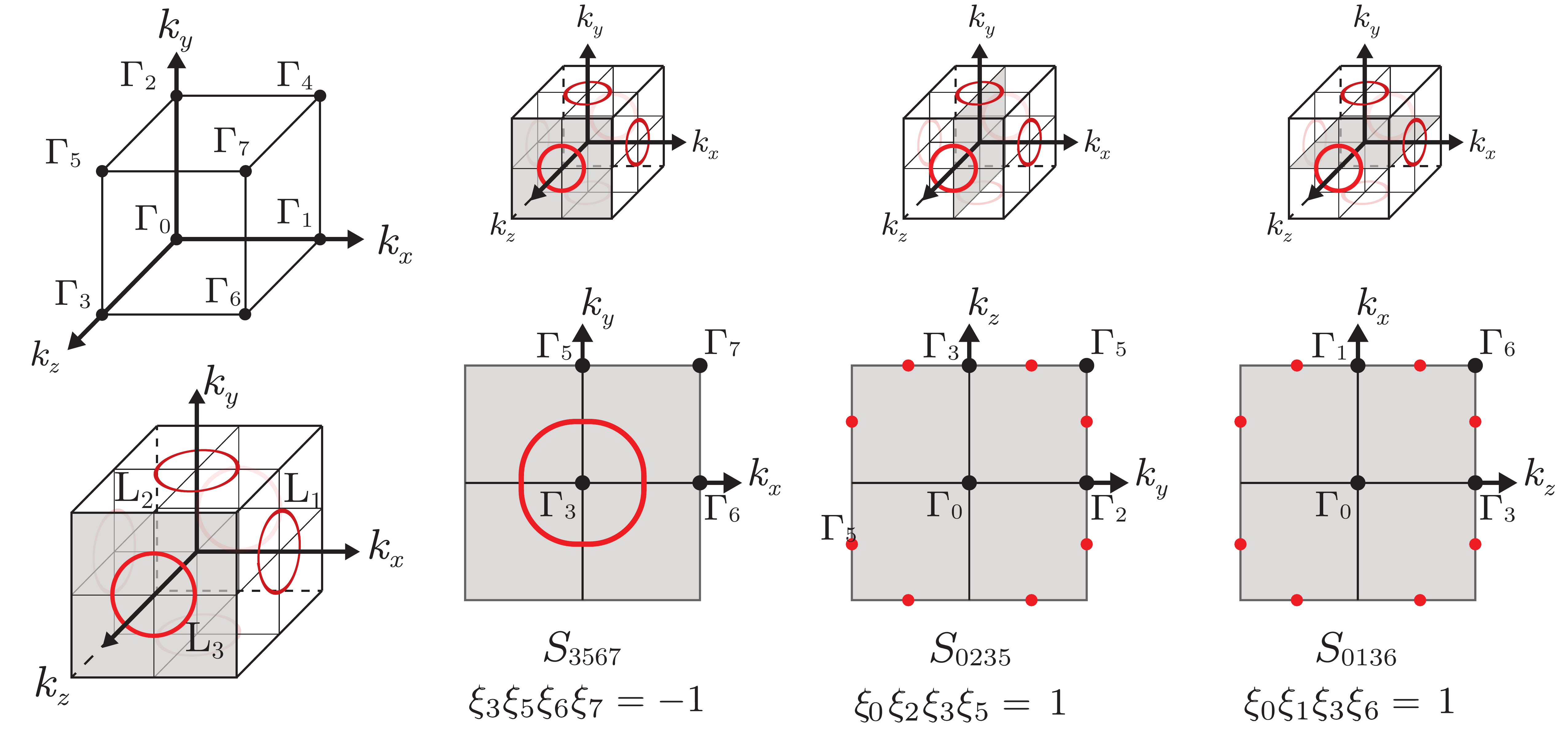}
\caption{\label{sfig3}
Dirac line nodes (DLNs) and invariant planes in the BZ of Cu$_3$NZn. Red circles represent the DLNs. The grey--shaded planes illustrate  the invariant planes that the DLN at $\G_3$ point ($L_3$) intersects, one of which ($S_{3567}$) is nontrivial and the other two ($S_{0235}$ and $S_{0136}$) are trivial. The intersecting position is depicted by red scheme on the planes.
}
\end{figure*}
First-principles calculations show that three distinct DLNs appear in Cu$_3$NZn, contained in the boundary planes of the BZ as shown in Fig.\ \ref{sfig3}. For convenience, we will refer to the DLN near the $\G_i$ as $L_i$, where $i =1, 2, 3$. The DLNs in Cu$_3$NZn form in a manner that is consistent with the topological constraint, imposed by $(\nu_0;\nu_1\nu_2\nu_3)=(1;111)$. To show this, we first consider an $L_3$ that encloses $\G_3$ in the BZ. As shown in Fig.\ \ref{sfig3}, $L_3$ intersects three invariant surfaces, referred to as $S_{3567}$, $S_{0235}$, and $S_{0136}$.  The $\ztwo$ topological invariants $\ztwos = (1;111)$  dictate that  $S_{3567}$ is nontrivial, and $S_{0235}$ and $S_{0136}$ are trivial, so that the nontrivial $S_{3567}$ will be pierced by an odd number of DLNs, while the trivial $S_{0136}$ and $S_{0235}$ planes will be pierced by an even (including zero) number of DLNs. It is clear from  Fig.\ \ref{sfig3} that the trivial $S_{0235}$ and $S_{0136}$ planes are pierced by two DLNs ($L_3$ and $L_1$ for $S_{0136}$, and $L_3$ and $L_2$ for $S_{0235}$), and is thus consistent with  $\x_a\x_b\x_c\x_d = 1$, where $(abcd) = (0235), (0136)$.
\begin{figure*}[tb!]
\includegraphics[width=0.85\textwidth]{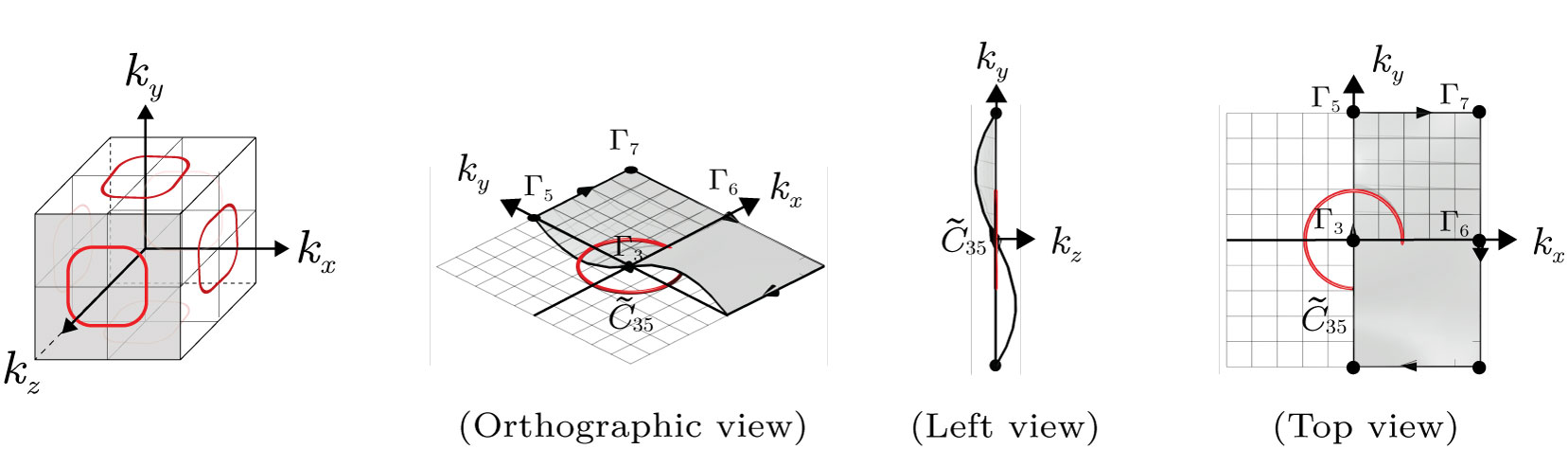}
\caption{\label{sfig4}
Invariant loop avoiding intersection with the DLN. The loop $c_{35}$ connecting $\G_3$ and $\G_5$ is bent down in the $-k_z$ direction, while its time--reversed partner $\bar c_{35}$ is bent up in the $k_z$ direction.  The corresponding interior surface $S_{3567}$ is threaded once by the DLN (illustrated by a red circle).
}
\end{figure*}
The nontrivial $S_{3567}$ plane also satisfies the corresponding topological constraint. To show this, we construct a $\mathcal{T}$--invariant loop $\tilde C_{35}$ that connects $\G_3$ and $\G_5$ avoiding the intersection with $L_3$ on the plane. The loop $c_{35}$ connecting $\G_3$ and $\G_5$ is bent down in the $-k_z$ direction, while its time--reversed partner $\bar c_{35}$ is bent up in the $k_z$ direction, as shown in Fig.\ \ref{sfig4}. From the figure, it is clear that the invariant surface $S_{3567}$, containing $\tilde C_{35}$ is pierced once by $L_3$ in its half plane, thus satisfying the topological constraint imposed by $\xi_3\xi_5\xi_6\xi_7 = -1$. Therefore, the appearance of $L_3$ satisfies the topological constraint. For the other inequivalent DLNs $L_1$ and $L_2$, the cubic symmetry of Cu$_3$NZn allows us to apply the same argument  for $L_1$ and $L_ 2$ to conclude that all the DLNs in Cu$_3$NZn obey the topological constraints characterized by $\ztwo$ topological invariants  $(\nu_0;\nu_1\nu_2\nu_3)=(1;111)$.

\subsection{Band structures beyond Cu$_3$NZn and Cu$_3$NPd}

\begin{figure*}[tb!]
\includegraphics[width=0.9\textwidth]{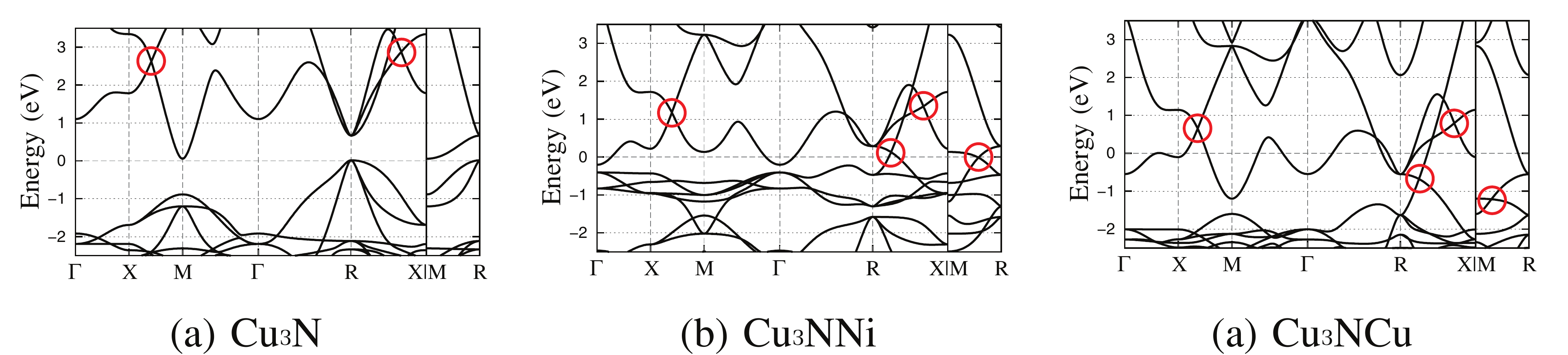}
\caption{\label{sfig5}
Band structures of Cu$_3$NX, with X=\{V$_0$, Ni, Zn\}. (a) Cu$_3$N. (b) Cu$_3$NNi, and (c) Cu$_3$NCu. The Dirac nodes are indicated by red circles.
}
\end{figure*}
In this section we extend our discussion of the material realization of DLN semimetals. We demonstrate that DLNs can occur in Cu$_3$N by doping a more general class of non--magnetic $3d$ and $4d$ transition metals (TMs) atoms in the $X$, $XI$, and $XII$ groups of the periodic table, and that Cu$_3$NPd$_x$ can realize DLNs in a low--doping concentration $x$ ($x < 1$). For this purpose, we first present the band structures of Cu$_3$NX, with X=\{Ni, Cu\} in Fig.\ \ref{sfig5}. These bands structures, including those of Cu$_3$NZn and Cu$_3$NPd are more or less similar, when considering the position--shift of the Fermi energy due to the electron valence of the TM dopants. The Fermi energy positions one band higher in the group $XII$ TM--doped case, comparing to that of the group $X$ TM--doped cases, due to two more valence electrons in the group $XII$ TMs than in the group $X$ TMs. Comparing to the band structure of Cu$_3$N in Fig.\ \ref{sfig5}(a), it is clear that the group $XII$ TMs provide two electrons per unit cell of Cu$_3$N, and thus move the preexisting DLNs near the Fermi energy. This indicates that a high doping concentration $x$ should be essential in the realization of the DLN near the Fermi energy in the group $XII$--doped systems.

\begin{figure*}[tb!]
\includegraphics[width=0.9\textwidth]{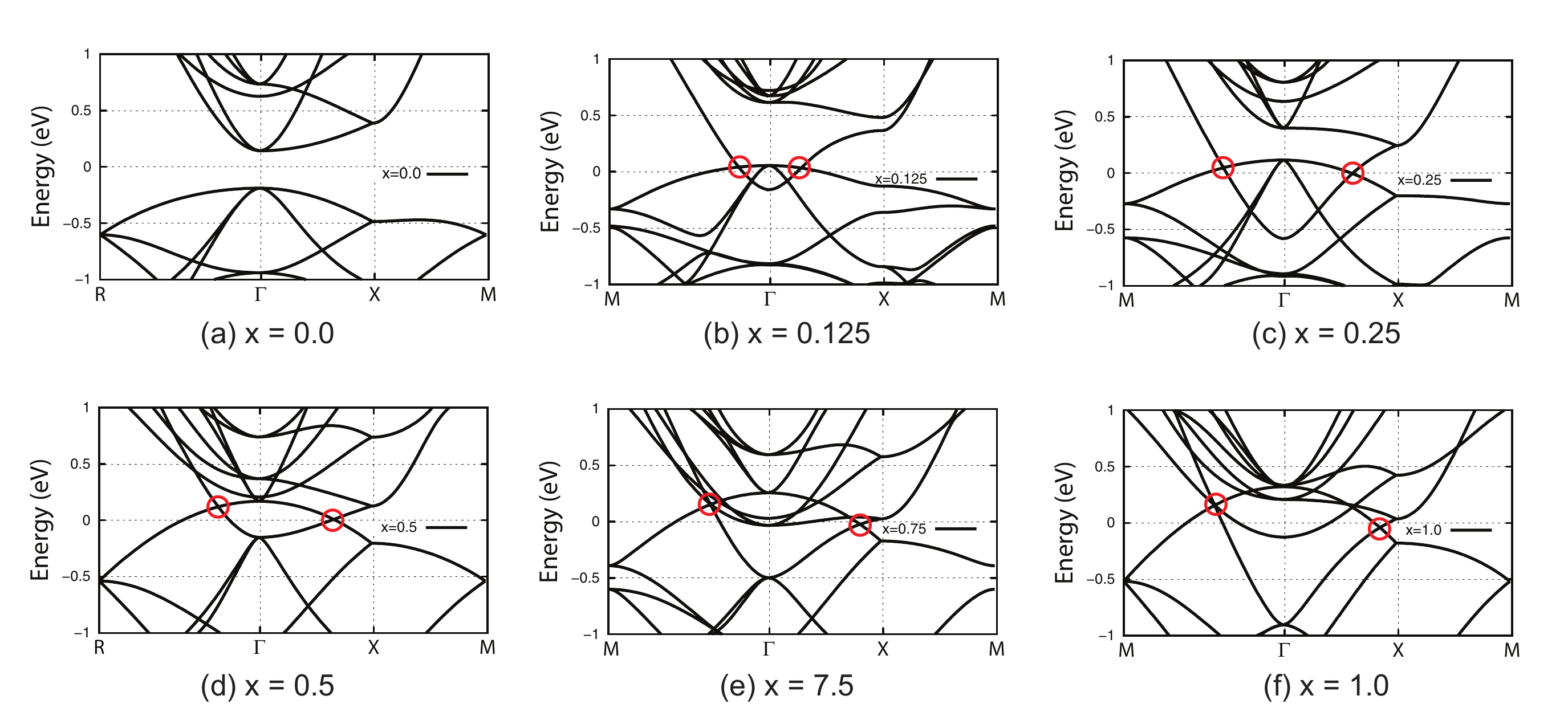}
\caption{\label{sfig6}
Band structures of Cu$_3$NPd$_x$. (a) x = 0.0, (b) x = 0.125, (c) x = 0.25, (d) x = 0.5, (e) x = 0.75, (f) x = 1.0. The Dirac nodes are indicated by red circles.
 }
\end{figure*}

However, in the group $X$--doped systems, such as Cu$_3$NPd, and Cu$_3$NNi [see Fig.\ \ref{sfig5}(b)], these high concentrations $(x \sim 1)$ are unnecessary. Under the doping of the group $X$ TMs, the Fermi level remains in the same region as where it was in Cu$_3$N, and instead doping drives the band--inversion. Therefore, DLNs start to appear near the Fermi energy in a lower concentration regime.  To demonstrate this, we calculate the band structures of Cu$_3$NPd$_x$, with  x = 0.0, 0.125, 0.25, 0.5, 0.75 and 1.0 by varying the number of Pd atoms in a  2$\times$2$\times$2 supercell of Cu$_3$N. The results, presented in Fig \ref{sfig6}, show that the conduction and valence bands are already inverted even at $x=0.125$, and that the degree of the band inversion, as well as the size of the DLN, increase with increasing doping concentration $x$.

\begin{figure*}[tb!]
\includegraphics[width=0.9\textwidth]{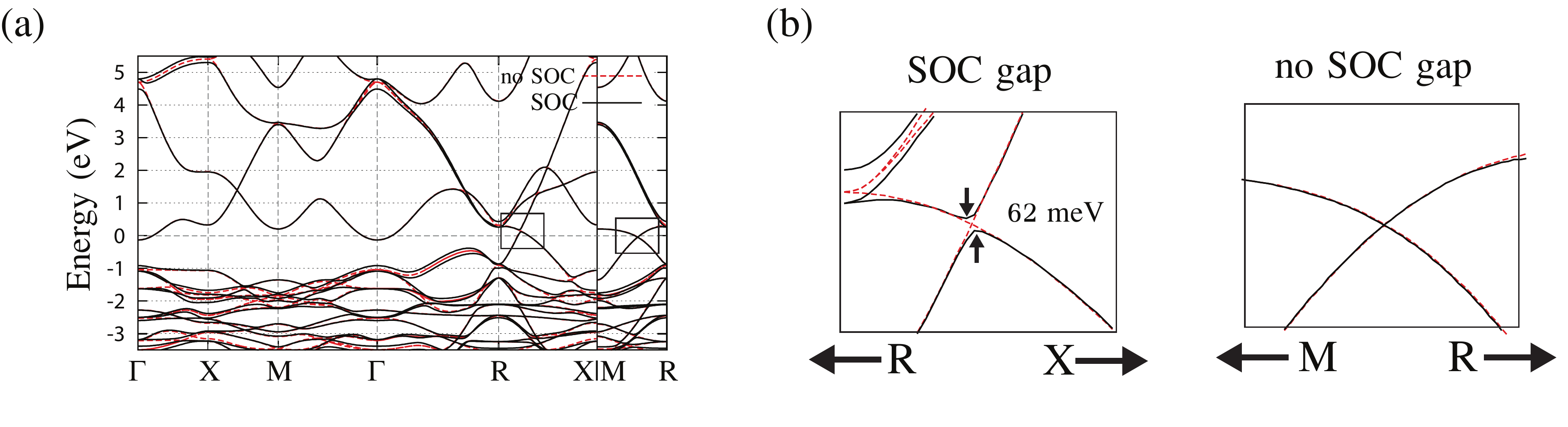}
\caption{\label{sfig7}
Band structures of Cu$_3$NPd with and without spin--orbit interaction. The bands in the rectangles in (a) are magnified in (b).  
}
\end{figure*}
Lastly, we consider the spin--orbit interaction in Cu$_3$NPd. In Fig.\ \ref{sfig7}, we present the band structures of Cu$_3$NPd calculated with and without spin--orbit coupling (SOC). It is clear that the SOC induces a sizable gap in the DLNs up to $\sim$ 62 meV along the $R$--$X$ high--symmetry line in the BZ. However, the SOC cannot completely gap out the entire DLN, as a single nodal point on the $R$--$X$ line additionally protected by a $C_4$ rotational symmetry.  The states forming the nodal point have distinct eigenvalues of the rotational symmetry operation, and thus retain the degeneracy even in the presence of the SOC. This is in fact one of the mechanisms which stabilizes three--dimensional Dirac (3D) semimetals \cite{Steve12p140405, Wang12p195320, Wang13p125427}. Therefore one might expect the 3D Dirac semimetal phase to be present in Cu$_3$NPd. However, considering that the SOC mainly comes from the $4d$ orbitals of Pd, and that the calculated SOC gap is estimated based on a somewhat extreme condition that Pd atoms are fully doped (one per unitcell of Cu$_3$N), it is more likely that the DLN semimetal should persist in a wide range of the Pd--doping concentration, and should be especially robust in the low--concentration regime. In order to strengthen (weaken) the SOC, one can substitute the dopants from $4d$ TMs to $5d$ ($3d$)  TMs such as Pt ($3d$ Ni). One can even dope with magnetic TMs to explore the effect of time--reversal--symmetry--breaking in the parent DLN semimetals.

\bibliography{ref}

\begin{thebibliography}{41}%
\makeatletter
\providecommand \@ifxundefined [1]{%
 \@ifx{#1\undefined}
}%
\providecommand \@ifnum [1]{%
 \ifnum #1\expandafter \@firstoftwo
 \else \expandafter \@secondoftwo
 \fi
}%
\providecommand \@ifx [1]{%
 \ifx #1\expandafter \@firstoftwo
 \else \expandafter \@secondoftwo
 \fi
}%
\providecommand \natexlab [1]{#1}%
\providecommand \enquote  [1]{``#1''}%
\providecommand \bibnamefont  [1]{#1}%
\providecommand \bibfnamefont [1]{#1}%
\providecommand \citenamefont [1]{#1}%
\providecommand \href@noop [0]{\@secondoftwo}%
\providecommand \href [0]{\begingroup \@sanitize@url \@href}%
\providecommand \@href[1]{\@@startlink{#1}\@@href}%
\providecommand \@@href[1]{\endgroup#1\@@endlink}%
\providecommand \@sanitize@url [0]{\catcode `\\12\catcode `\$12\catcode
  `\&12\catcode `\#12\catcode `\^12\catcode `\_12\catcode `\%12\relax}%
\providecommand \@@startlink[1]{}%
\providecommand \@@endlink[0]{}%
\providecommand \url  [0]{\begingroup\@sanitize@url \@url }%
\providecommand \@url [1]{\endgroup\@href {#1}{\urlprefix }}%
\providecommand \urlprefix  [0]{URL }%
\providecommand \Eprint [0]{\href }%
\providecommand \doibase [0]{http://dx.doi.org/}%
\providecommand \selectlanguage [0]{\@gobble}%
\providecommand \bibinfo  [0]{\@secondoftwo}%
\providecommand \bibfield  [0]{\@secondoftwo}%
\providecommand \translation [1]{[#1]}%
\providecommand \BibitemOpen [0]{}%
\providecommand \bibitemStop [0]{}%
\providecommand \bibitemNoStop [0]{.\EOS\space}%
\providecommand \EOS [0]{\spacefactor3000\relax}%
\providecommand \BibitemShut  [1]{\csname bibitem#1\endcsname}%
\let\auto@bib@innerbib\@empty
\bibitem [{\citenamefont {Hasan}\ and\ \citenamefont
  {Kane}(2010)}]{Hasan10p3045}%
  \BibitemOpen
  \bibfield  {author} {\bibinfo {author} {\bibfnamefont {M.~Z.}\ \bibnamefont
  {Hasan}}\ and\ \bibinfo {author} {\bibfnamefont {C.~L.}\ \bibnamefont
  {Kane}},\ }\href {\doibase 10.1103/RevModPhys.82.3045} {\bibfield  {journal}
  {\bibinfo  {journal} {Rev. Mod. Phys.}\ }\textbf {\bibinfo {volume} {82}},\
  \bibinfo {pages} {3045} (\bibinfo {year} {2010})}\BibitemShut {NoStop}%
\bibitem [{\citenamefont {Qi}\ and\ \citenamefont {Zhang}(2011)}]{Qi11p1057}%
  \BibitemOpen
  \bibfield  {author} {\bibinfo {author} {\bibfnamefont {X.-L.}\ \bibnamefont
  {Qi}}\ and\ \bibinfo {author} {\bibfnamefont {S.-C.}\ \bibnamefont {Zhang}},\
  }\href {\doibase 10.1103/RevModPhys.83.1057} {\bibfield  {journal} {\bibinfo
  {journal} {Rev. Mod. Phys.}\ }\textbf {\bibinfo {volume} {83}},\ \bibinfo
  {pages} {1057} (\bibinfo {year} {2011})}\BibitemShut {NoStop}%
\bibitem [{\citenamefont {Fu}(2011)}]{Fu11p106802}%
  \BibitemOpen
  \bibfield  {author} {\bibinfo {author} {\bibfnamefont {L.}~\bibnamefont
  {Fu}},\ }\href {\doibase 10.1103/PhysRevLett.106.106802} {\bibfield
  {journal} {\bibinfo  {journal} {Phys. Rev. Lett.}\ }\textbf {\bibinfo
  {volume} {106}},\ \bibinfo {pages} {106802} (\bibinfo {year}
  {2011})}\BibitemShut {NoStop}%
\bibitem [{\citenamefont {Wan}\ \emph {et~al.}(2011)\citenamefont {Wan},
  \citenamefont {Turner}, \citenamefont {Vishwanath},\ and\ \citenamefont
  {Savrasov}}]{Wan11p205101}%
  \BibitemOpen
  \bibfield  {author} {\bibinfo {author} {\bibfnamefont {X.}~\bibnamefont
  {Wan}}, \bibinfo {author} {\bibfnamefont {A.~M.}\ \bibnamefont {Turner}},
  \bibinfo {author} {\bibfnamefont {A.}~\bibnamefont {Vishwanath}}, \ and\
  \bibinfo {author} {\bibfnamefont {S.~Y.}\ \bibnamefont {Savrasov}},\ }\href
  {\doibase 10.1103/PhysRevB.83.205101} {\bibfield  {journal} {\bibinfo
  {journal} {Phys. Rev. B}\ }\textbf {\bibinfo {volume} {83}},\ \bibinfo
  {pages} {205101} (\bibinfo {year} {2011})}\BibitemShut {NoStop}%
\bibitem [{\citenamefont {Young}\ \emph {et~al.}(2012)\citenamefont {Young},
  \citenamefont {Zaheer}, \citenamefont {Teo}, \citenamefont {Kane},
  \citenamefont {Mele},\ and\ \citenamefont {Rappe}}]{Steve12p140405}%
  \BibitemOpen
  \bibfield  {author} {\bibinfo {author} {\bibfnamefont {S.~M.}\ \bibnamefont
  {Young}}, \bibinfo {author} {\bibfnamefont {S.}~\bibnamefont {Zaheer}},
  \bibinfo {author} {\bibfnamefont {J.~C.~Y.}\ \bibnamefont {Teo}}, \bibinfo
  {author} {\bibfnamefont {C.~L.}\ \bibnamefont {Kane}}, \bibinfo {author}
  {\bibfnamefont {E.~J.}\ \bibnamefont {Mele}}, \ and\ \bibinfo {author}
  {\bibfnamefont {A.~M.}\ \bibnamefont {Rappe}},\ }\href {\doibase
  10.1103/PhysRevLett.108.140405} {\bibfield  {journal} {\bibinfo  {journal}
  {Phys. Rev. Lett.}\ }\textbf {\bibinfo {volume} {108}},\ \bibinfo {pages}
  {140405} (\bibinfo {year} {2012})}\BibitemShut {NoStop}%
\bibitem [{\citenamefont {Steinberg}\ \emph {et~al.}(2014)\citenamefont
  {Steinberg}, \citenamefont {Young}, \citenamefont {Zaheer}, \citenamefont
  {Kane}, \citenamefont {Mele},\ and\ \citenamefont
  {Rappe}}]{Steinberg14p036403}%
  \BibitemOpen
  \bibfield  {author} {\bibinfo {author} {\bibfnamefont {J.~A.}\ \bibnamefont
  {Steinberg}}, \bibinfo {author} {\bibfnamefont {S.~M.}\ \bibnamefont
  {Young}}, \bibinfo {author} {\bibfnamefont {S.}~\bibnamefont {Zaheer}},
  \bibinfo {author} {\bibfnamefont {C.~L.}\ \bibnamefont {Kane}}, \bibinfo
  {author} {\bibfnamefont {E.~J.}\ \bibnamefont {Mele}}, \ and\ \bibinfo
  {author} {\bibfnamefont {A.~M.}\ \bibnamefont {Rappe}},\ }\href {\doibase
  10.1103/PhysRevLett.112.036403} {\bibfield  {journal} {\bibinfo  {journal}
  {Phys. Rev. Lett.}\ }\textbf {\bibinfo {volume} {112}},\ \bibinfo {pages}
  {036403} (\bibinfo {year} {2014})}\BibitemShut {NoStop}%
\bibitem [{\citenamefont {Alexandradinata}\ \emph {et~al.}(2014)\citenamefont
  {Alexandradinata}, \citenamefont {Fang}, \citenamefont {Gilbert},\ and\
  \citenamefont {Bernevig}}]{Alexandradinata14p116403}%
  \BibitemOpen
  \bibfield  {author} {\bibinfo {author} {\bibfnamefont {A.}~\bibnamefont
  {Alexandradinata}}, \bibinfo {author} {\bibfnamefont {C.}~\bibnamefont
  {Fang}}, \bibinfo {author} {\bibfnamefont {M.~J.}\ \bibnamefont {Gilbert}}, \
  and\ \bibinfo {author} {\bibfnamefont {B.~A.}\ \bibnamefont {Bernevig}},\
  }\href {\doibase 10.1103/PhysRevLett.113.116403} {\bibfield  {journal}
  {\bibinfo  {journal} {Phys. Rev. Lett.}\ }\textbf {\bibinfo {volume} {113}},\
  \bibinfo {pages} {116403} (\bibinfo {year} {2014})}\BibitemShut {NoStop}%
\bibitem [{\citenamefont {Castro~Neto}\ \emph {et~al.}(2009)\citenamefont
  {Castro~Neto}, \citenamefont {Guinea}, \citenamefont {Peres}, \citenamefont
  {Novoselov},\ and\ \citenamefont {Geim}}]{Neto09p109}%
  \BibitemOpen
  \bibfield  {author} {\bibinfo {author} {\bibfnamefont {A.~H.}\ \bibnamefont
  {Castro~Neto}}, \bibinfo {author} {\bibfnamefont {F.}~\bibnamefont {Guinea}},
  \bibinfo {author} {\bibfnamefont {N.~M.~R.}\ \bibnamefont {Peres}}, \bibinfo
  {author} {\bibfnamefont {K.~S.}\ \bibnamefont {Novoselov}}, \ and\ \bibinfo
  {author} {\bibfnamefont {A.~K.}\ \bibnamefont {Geim}},\ }\href {\doibase
  10.1103/RevModPhys.81.109} {\bibfield  {journal} {\bibinfo  {journal} {Rev.
  Mod. Phys.}\ }\textbf {\bibinfo {volume} {81}},\ \bibinfo {pages} {109}
  (\bibinfo {year} {2009})}\BibitemShut {NoStop}%
\bibitem [{\citenamefont {Weng}\ \emph {et~al.}(2014)\citenamefont {Weng},
  \citenamefont {Liang}, \citenamefont {Xu}, \citenamefont {Rui}, \citenamefont
  {Fang}, \citenamefont {Dai},\ and\ \citenamefont {Kawazoe}}]{Weng14p1}%
  \BibitemOpen
  \bibfield  {author} {\bibinfo {author} {\bibfnamefont {H.}~\bibnamefont
  {Weng}}, \bibinfo {author} {\bibfnamefont {Y.}~\bibnamefont {Liang}},
  \bibinfo {author} {\bibfnamefont {Q.}~\bibnamefont {Xu}}, \bibinfo {author}
  {\bibfnamefont {Y.}~\bibnamefont {Rui}}, \bibinfo {author} {\bibfnamefont
  {Z.}~\bibnamefont {Fang}}, \bibinfo {author} {\bibfnamefont {X.}~\bibnamefont
  {Dai}}, \ and\ \bibinfo {author} {\bibfnamefont {Y.}~\bibnamefont
  {Kawazoe}},\ }\href@noop {} {\enquote {\bibinfo {title} {{Topological
  Node-Line Semimetal in Three Dimensional Graphene Networks}},}\ } (\bibinfo
  {year} {2014}),\ \Eprint {http://arxiv.org/abs/arXiv:1411.2175}
  {arXiv:1411.2175} \BibitemShut {NoStop}%
\bibitem [{\citenamefont {Volovik}(2011)}]{Volovik11p1}%
  \BibitemOpen
  \bibfield  {author} {\bibinfo {author} {\bibfnamefont {G.~E.}\ \bibnamefont
  {Volovik}},\ }\href@noop {} {\enquote {\bibinfo {title} {Flat band in
  topological matter: possible route to room-temperature superconductivity},}\
  } (\bibinfo {year} {2011}),\ \Eprint {http://arxiv.org/abs/arXiv:1110.4469}
  {arXiv:1110.4469} \BibitemShut {NoStop}%
\bibitem [{\citenamefont {Kim}\ \emph {et~al.}(2014)\citenamefont {Kim},
  \citenamefont {Chen},\ and\ \citenamefont {Kee}}]{Kim14p1}%
  \BibitemOpen
  \bibfield  {author} {\bibinfo {author} {\bibfnamefont {H.-S.}\ \bibnamefont
  {Kim}}, \bibinfo {author} {\bibfnamefont {Y.}~\bibnamefont {Chen}}, \ and\
  \bibinfo {author} {\bibfnamefont {H.-Y.}\ \bibnamefont {Kee}},\ }\href@noop
  {} {\enquote {\bibinfo {title} {{Surface States of Perovskite Iridates
  AIrO$_3$; Signatures of Topological Crystalline Metal with Nontrivial
  $\mathbb{Z}_2$ Index}},}\ } (\bibinfo {year} {2014}),\ \Eprint
  {http://arxiv.org/abs/arXiv:1411.1406} {arXiv:1411.1406} \BibitemShut
  {NoStop}%
\bibitem [{\citenamefont {Chen}\ \emph {et~al.}(2015)\citenamefont {Chen},
  \citenamefont {Lu},\ and\ \citenamefont {Kee}}]{Chen15p1}%
  \BibitemOpen
  \bibfield  {author} {\bibinfo {author} {\bibfnamefont {Y.}~\bibnamefont
  {Chen}}, \bibinfo {author} {\bibfnamefont {Y.-M.}\ \bibnamefont {Lu}}, \ and\
  \bibinfo {author} {\bibfnamefont {H.-Y.}\ \bibnamefont {Kee}},\ }\href@noop
  {} {\bibfield  {journal} {\bibinfo  {journal} {Nat. Commun.}\ }\textbf
  {\bibinfo {volume} {6}},\ \bibinfo {pages} {6593} (\bibinfo {year}
  {2015})}\BibitemShut {NoStop}%
\bibitem [{\citenamefont {Weng}\ \emph {et~al.}(2015)\citenamefont {Weng},
  \citenamefont {Fang}, \citenamefont {Fang}, \citenamefont {Bernevig},\ and\
  \citenamefont {Dai}}]{Weng15p011029}%
  \BibitemOpen
  \bibfield  {author} {\bibinfo {author} {\bibfnamefont {H.}~\bibnamefont
  {Weng}}, \bibinfo {author} {\bibfnamefont {C.}~\bibnamefont {Fang}}, \bibinfo
  {author} {\bibfnamefont {Z.}~\bibnamefont {Fang}}, \bibinfo {author}
  {\bibfnamefont {B.~A.}\ \bibnamefont {Bernevig}}, \ and\ \bibinfo {author}
  {\bibfnamefont {X.}~\bibnamefont {Dai}},\ }\href {\doibase
  10.1103/PhysRevX.5.011029} {\bibfield  {journal} {\bibinfo  {journal} {Phys.
  Rev. X}\ }\textbf {\bibinfo {volume} {5}},\ \bibinfo {pages} {011029}
  (\bibinfo {year} {2015})}\BibitemShut {NoStop}%
\bibitem [{Note1()}]{Note1}%
  \BibitemOpen
  \bibinfo {note} {Since $[\protect \mathcal {H}({\protect \bf k}),\protect
  \mathcal {P}\protect \mathcal {T}]=0$, this $\protect \mathcal {Z}_2$
  invariant can also be understood as characterizing one parameter families of
  Hamiltonians in class AI \cite {Teo10p115120}}\BibitemShut {NoStop}%
\bibitem [{\citenamefont {Fu}\ \emph {et~al.}(2007)\citenamefont {Fu},
  \citenamefont {Kane},\ and\ \citenamefont {Mele}}]{Fu07p106803}%
  \BibitemOpen
  \bibfield  {author} {\bibinfo {author} {\bibfnamefont {L.}~\bibnamefont
  {Fu}}, \bibinfo {author} {\bibfnamefont {C.~L.}\ \bibnamefont {Kane}}, \ and\
  \bibinfo {author} {\bibfnamefont {E.~J.}\ \bibnamefont {Mele}},\ }\href
  {\doibase 10.1103/PhysRevLett.98.106803} {\bibfield  {journal} {\bibinfo
  {journal} {Phys. Rev. Lett.}\ }\textbf {\bibinfo {volume} {98}},\ \bibinfo
  {pages} {106803} (\bibinfo {year} {2007})}\BibitemShut {NoStop}%
\bibitem [{\citenamefont {Fu}\ and\ \citenamefont {Kane}(2007)}]{Fu07p045302}%
  \BibitemOpen
  \bibfield  {author} {\bibinfo {author} {\bibfnamefont {L.}~\bibnamefont
  {Fu}}\ and\ \bibinfo {author} {\bibfnamefont {C.~L.}\ \bibnamefont {Kane}},\
  }\href {\doibase 10.1103/PhysRevB.76.045302} {\bibfield  {journal} {\bibinfo
  {journal} {Phys. Rev. B}\ }\textbf {\bibinfo {volume} {76}},\ \bibinfo
  {pages} {045302} (\bibinfo {year} {2007})}\BibitemShut {NoStop}%
\bibitem [{\citenamefont {Juza}\ and\ \citenamefont {Hahn}(1938)}]{Juza38p282}%
  \BibitemOpen
  \bibfield  {author} {\bibinfo {author} {\bibfnamefont {R.}~\bibnamefont
  {Juza}}\ and\ \bibinfo {author} {\bibfnamefont {H.}~\bibnamefont {Hahn}},\
  }\href {\doibase 10.1002/zaac.19382390307} {\bibfield  {journal} {\bibinfo
  {journal} {Zeitschrift für anorganische und allgemeine Chemie}\ }\textbf
  {\bibinfo {volume} {239}},\ \bibinfo {pages} {282} (\bibinfo {year}
  {1938})}\BibitemShut {NoStop}%
\bibitem [{\citenamefont {Hadian}\ \emph {et~al.}(2012)\citenamefont {Hadian},
  \citenamefont {Rahmati}, \citenamefont {Movla},\ and\ \citenamefont
  {Khaksar}}]{Hadian12p1067}%
  \BibitemOpen
  \bibfield  {author} {\bibinfo {author} {\bibfnamefont {F.}~\bibnamefont
  {Hadian}}, \bibinfo {author} {\bibfnamefont {A.}~\bibnamefont {Rahmati}},
  \bibinfo {author} {\bibfnamefont {H.}~\bibnamefont {Movla}}, \ and\ \bibinfo
  {author} {\bibfnamefont {M.}~\bibnamefont {Khaksar}},\ }\href {\doibase
  http://dx.doi.org/10.1016/j.vacuum.2011.09.001} {\bibfield  {journal}
  {\bibinfo  {journal} {Vacuum}\ }\textbf {\bibinfo {volume} {86}},\ \bibinfo
  {pages} {1067 } (\bibinfo {year} {2012})}\BibitemShut {NoStop}%
\bibitem [{\citenamefont {Gulo}\ \emph {et~al.}(2004)\citenamefont {Gulo},
  \citenamefont {Simon}, \citenamefont {Köhler},\ and\ \citenamefont
  {Kremer}}]{Gulo04p2032}%
  \BibitemOpen
  \bibfield  {author} {\bibinfo {author} {\bibfnamefont {F.}~\bibnamefont
  {Gulo}}, \bibinfo {author} {\bibfnamefont {A.}~\bibnamefont {Simon}},
  \bibinfo {author} {\bibfnamefont {J.}~\bibnamefont {Köhler}}, \ and\
  \bibinfo {author} {\bibfnamefont {R.~K.}\ \bibnamefont {Kremer}},\ }\href
  {\doibase 10.1002/anie.200353424} {\bibfield  {journal} {\bibinfo  {journal}
  {Angewandte Chemie International Edition}\ }\textbf {\bibinfo {volume}
  {43}},\ \bibinfo {pages} {2032} (\bibinfo {year} {2004})}\BibitemShut
  {NoStop}%
\bibitem [{\citenamefont {Hou}(2008)}]{Hou08p1651}%
  \BibitemOpen
  \bibfield  {author} {\bibinfo {author} {\bibfnamefont {Z.}~\bibnamefont
  {Hou}},\ }\href {\doibase
  http://dx.doi.org/10.1016/j.solidstatesciences.2008.02.013} {\bibfield
  {journal} {\bibinfo  {journal} {Solid State Sciences}\ }\textbf {\bibinfo
  {volume} {10}},\ \bibinfo {pages} {1651 } (\bibinfo {year}
  {2008})}\BibitemShut {NoStop}%
\bibitem [{\citenamefont {Moreno-Armenta}\ \emph {et~al.}(2007)\citenamefont
  {Moreno-Armenta}, \citenamefont {Pérez},\ and\ \citenamefont
  {Takeuchi}}]{MorenoArmenta07p166}%
  \BibitemOpen
  \bibfield  {author} {\bibinfo {author} {\bibfnamefont {M.~G.}\ \bibnamefont
  {Moreno-Armenta}}, \bibinfo {author} {\bibfnamefont {W.~L.}\ \bibnamefont
  {Pérez}}, \ and\ \bibinfo {author} {\bibfnamefont {N.}~\bibnamefont
  {Takeuchi}},\ }\href {\doibase
  http://dx.doi.org/10.1016/j.solidstatesciences.2006.12.002} {\bibfield
  {journal} {\bibinfo  {journal} {Solid State Sciences}\ }\textbf {\bibinfo
  {volume} {9}},\ \bibinfo {pages} {166 } (\bibinfo {year} {2007})}\BibitemShut
  {NoStop}%
\bibitem [{\citenamefont {Zachwieja}\ and\ \citenamefont
  {Jacobs}(1990)}]{Zachwieja90p175}%
  \BibitemOpen
  \bibfield  {author} {\bibinfo {author} {\bibfnamefont {U.}~\bibnamefont
  {Zachwieja}}\ and\ \bibinfo {author} {\bibfnamefont {H.}~\bibnamefont
  {Jacobs}},\ }\href {\doibase http://dx.doi.org/10.1016/0022-5088(90)90327-G}
  {\bibfield  {journal} {\bibinfo  {journal} {Journal of the Less Common
  Metals}\ }\textbf {\bibinfo {volume} {161}},\ \bibinfo {pages} {175 }
  (\bibinfo {year} {1990})}\BibitemShut {NoStop}%
\bibitem [{\citenamefont {Hahn}\ and\ \citenamefont
  {Weber}(1996)}]{Hahn96p12684}%
  \BibitemOpen
  \bibfield  {author} {\bibinfo {author} {\bibfnamefont {U.}~\bibnamefont
  {Hahn}}\ and\ \bibinfo {author} {\bibfnamefont {W.}~\bibnamefont {Weber}},\
  }\href {\doibase 10.1103/PhysRevB.53.12684} {\bibfield  {journal} {\bibinfo
  {journal} {Phys. Rev. B}\ }\textbf {\bibinfo {volume} {53}},\ \bibinfo
  {pages} {12684} (\bibinfo {year} {1996})}\BibitemShut {NoStop}%
\bibitem [{\citenamefont {Ji}\ \emph {et~al.}(2006)\citenamefont {Ji},
  \citenamefont {Li},\ and\ \citenamefont {Cao}}]{Ji06p252120}%
  \BibitemOpen
  \bibfield  {author} {\bibinfo {author} {\bibfnamefont {A.}~\bibnamefont
  {Ji}}, \bibinfo {author} {\bibfnamefont {C.}~\bibnamefont {Li}}, \ and\
  \bibinfo {author} {\bibfnamefont {Z.}~\bibnamefont {Cao}},\ }\href {\doibase
  http://dx.doi.org/10.1063/1.2422882} {\bibfield  {journal} {\bibinfo
  {journal} {Applied Physics Letters}\ }\textbf {\bibinfo {volume} {89}},\
  \bibinfo {eid} {252120} (\bibinfo {year} {2006})}\BibitemShut {NoStop}%
\bibitem [{\citenamefont {Sieberer}\ \emph {et~al.}(2006)\citenamefont
  {Sieberer}, \citenamefont {Khmelevskyi},\ and\ \citenamefont
  {Mohn}}]{Sieberer06p014416}%
  \BibitemOpen
  \bibfield  {author} {\bibinfo {author} {\bibfnamefont {M.}~\bibnamefont
  {Sieberer}}, \bibinfo {author} {\bibfnamefont {S.}~\bibnamefont
  {Khmelevskyi}}, \ and\ \bibinfo {author} {\bibfnamefont {P.}~\bibnamefont
  {Mohn}},\ }\href {\doibase 10.1103/PhysRevB.74.014416} {\bibfield  {journal}
  {\bibinfo  {journal} {Phys. Rev. B}\ }\textbf {\bibinfo {volume} {74}},\
  \bibinfo {pages} {014416} (\bibinfo {year} {2006})}\BibitemShut {NoStop}%
\bibitem [{\citenamefont {Moreno-Armenta}\ \emph {et~al.}(2004)\citenamefont
  {Moreno-Armenta}, \citenamefont {Martı́nez-Ruiz},\ and\ \citenamefont
  {Takeuchi}}]{MorenoArmenta04p9}%
  \BibitemOpen
  \bibfield  {author} {\bibinfo {author} {\bibfnamefont {M.~G.}\ \bibnamefont
  {Moreno-Armenta}}, \bibinfo {author} {\bibfnamefont {A.}~\bibnamefont
  {Martı́nez-Ruiz}}, \ and\ \bibinfo {author} {\bibfnamefont
  {N.}~\bibnamefont {Takeuchi}},\ }\href {\doibase
  http://dx.doi.org/10.1016/j.solidstatesciences.2003.10.014} {\bibfield
  {journal} {\bibinfo  {journal} {Solid State Sciences}\ }\textbf {\bibinfo
  {volume} {6}},\ \bibinfo {pages} {9 } (\bibinfo {year} {2004})}\BibitemShut
  {NoStop}%
\bibitem [{\citenamefont {Fan}\ \emph {et~al.}(2007)\citenamefont {Fan},
  \citenamefont {Wu}, \citenamefont {Zhang}, \citenamefont {Li}, \citenamefont
  {Geng}, \citenamefont {Li},\ and\ \citenamefont {Yan}}]{Fan07p254}%
  \BibitemOpen
  \bibfield  {author} {\bibinfo {author} {\bibfnamefont {X.}~\bibnamefont
  {Fan}}, \bibinfo {author} {\bibfnamefont {Z.}~\bibnamefont {Wu}}, \bibinfo
  {author} {\bibfnamefont {G.}~\bibnamefont {Zhang}}, \bibinfo {author}
  {\bibfnamefont {C.}~\bibnamefont {Li}}, \bibinfo {author} {\bibfnamefont
  {B.}~\bibnamefont {Geng}}, \bibinfo {author} {\bibfnamefont {H.}~\bibnamefont
  {Li}}, \ and\ \bibinfo {author} {\bibfnamefont {P.}~\bibnamefont {Yan}},\
  }\href {\doibase http://dx.doi.org/10.1016/j.jallcom.2006.09.006} {\bibfield
  {journal} {\bibinfo  {journal} {Journal of Alloys and Compounds}\ }\textbf
  {\bibinfo {volume} {440}},\ \bibinfo {pages} {254 } (\bibinfo {year}
  {2007})}\BibitemShut {NoStop}%
\bibitem [{\citenamefont {Pierson}\ and\ \citenamefont
  {Horwat}(2008)}]{Pierson08p568}%
  \BibitemOpen
  \bibfield  {author} {\bibinfo {author} {\bibfnamefont {J.}~\bibnamefont
  {Pierson}}\ and\ \bibinfo {author} {\bibfnamefont {D.}~\bibnamefont
  {Horwat}},\ }\href {\doibase
  http://dx.doi.org/10.1016/j.scriptamat.2007.11.016} {\bibfield  {journal}
  {\bibinfo  {journal} {Scripta Materialia}\ }\textbf {\bibinfo {volume}
  {58}},\ \bibinfo {pages} {568 } (\bibinfo {year} {2008})}\BibitemShut
  {NoStop}%
\bibitem [{\citenamefont {Wu}\ \emph {et~al.}(2014)\citenamefont {Wu},
  \citenamefont {Chen}, \citenamefont {Gao}, \citenamefont {Zhang},
  \citenamefont {Yang}, \citenamefont {Yang}, \citenamefont {Li},\ and\
  \citenamefont {Huang}}]{Wu14p221}%
  \BibitemOpen
  \bibfield  {author} {\bibinfo {author} {\bibfnamefont {Z.}~\bibnamefont
  {Wu}}, \bibinfo {author} {\bibfnamefont {H.}~\bibnamefont {Chen}}, \bibinfo
  {author} {\bibfnamefont {N.}~\bibnamefont {Gao}}, \bibinfo {author}
  {\bibfnamefont {E.}~\bibnamefont {Zhang}}, \bibinfo {author} {\bibfnamefont
  {J.}~\bibnamefont {Yang}}, \bibinfo {author} {\bibfnamefont {T.}~\bibnamefont
  {Yang}}, \bibinfo {author} {\bibfnamefont {X.}~\bibnamefont {Li}}, \ and\
  \bibinfo {author} {\bibfnamefont {W.}~\bibnamefont {Huang}},\ }\href
  {\doibase http://dx.doi.org/10.1016/j.commatsci.2014.07.035} {\bibfield
  {journal} {\bibinfo  {journal} {Computational Materials Science}\ }\textbf
  {\bibinfo {volume} {95}},\ \bibinfo {pages} {221 } (\bibinfo {year}
  {2014})}\BibitemShut {NoStop}%
\bibitem [{\citenamefont {Cui}\ \emph {et~al.}(2012)\citenamefont {Cui},
  \citenamefont {Soon}, \citenamefont {Phillips}, \citenamefont {Zheng},
  \citenamefont {Liu}, \citenamefont {Delley}, \citenamefont {Ringer},\ and\
  \citenamefont {Stampfl}}]{Cui12p3138}%
  \BibitemOpen
  \bibfield  {author} {\bibinfo {author} {\bibfnamefont {X.}~\bibnamefont
  {Cui}}, \bibinfo {author} {\bibfnamefont {A.}~\bibnamefont {Soon}}, \bibinfo
  {author} {\bibfnamefont {A.}~\bibnamefont {Phillips}}, \bibinfo {author}
  {\bibfnamefont {R.}~\bibnamefont {Zheng}}, \bibinfo {author} {\bibfnamefont
  {Z.}~\bibnamefont {Liu}}, \bibinfo {author} {\bibfnamefont {B.}~\bibnamefont
  {Delley}}, \bibinfo {author} {\bibfnamefont {S.}~\bibnamefont {Ringer}}, \
  and\ \bibinfo {author} {\bibfnamefont {C.}~\bibnamefont {Stampfl}},\ }\href
  {\doibase http://dx.doi.org/10.1016/j.jmmm.2012.05.021} {\bibfield  {journal}
  {\bibinfo  {journal} {Journal of Magnetism and Magnetic Materials}\ }\textbf
  {\bibinfo {volume} {324}},\ \bibinfo {pages} {3138 } (\bibinfo {year}
  {2012})}\BibitemShut {NoStop}%
\bibitem [{\citenamefont {Perdew}\ \emph {et~al.}(1996)\citenamefont {Perdew},
  \citenamefont {Burke},\ and\ \citenamefont {Ernzerhof}}]{Perdew96p3865}%
  \BibitemOpen
  \bibfield  {author} {\bibinfo {author} {\bibfnamefont {J.~P.}\ \bibnamefont
  {Perdew}}, \bibinfo {author} {\bibfnamefont {K.}~\bibnamefont {Burke}}, \
  and\ \bibinfo {author} {\bibfnamefont {M.}~\bibnamefont {Ernzerhof}},\ }\href
  {\doibase 10.1103/PhysRevLett.77.3865} {\bibfield  {journal} {\bibinfo
  {journal} {Phys. Rev. Lett.}\ }\textbf {\bibinfo {volume} {77}},\ \bibinfo
  {pages} {3865} (\bibinfo {year} {1996})}\BibitemShut {NoStop}%
\bibitem [{\citenamefont {Giannozzi}\ \emph {et~al.}(2009)\citenamefont
  {Giannozzi} \emph {et~al.}}]{Giannozzi09p395502}%
  \BibitemOpen
  \bibfield  {author} {\bibinfo {author} {\bibfnamefont {P.}~\bibnamefont
  {Giannozzi}} \emph {et~al.},\ }\href@noop {} {\bibfield  {journal} {\bibinfo
  {journal} {Journal of Physics: Condensed Matter}\ }\textbf {\bibinfo {volume}
  {21}},\ \bibinfo {pages} {395502} (\bibinfo {year} {2009})}\BibitemShut
  {NoStop}%
\bibitem [{\citenamefont {Rappe}\ \emph {et~al.}(1990)\citenamefont {Rappe},
  \citenamefont {Rabe}, \citenamefont {Kaxiras},\ and\ \citenamefont
  {Joannopoulos}}]{Rappe90p1227}%
  \BibitemOpen
  \bibfield  {author} {\bibinfo {author} {\bibfnamefont {A.~M.}\ \bibnamefont
  {Rappe}}, \bibinfo {author} {\bibfnamefont {K.~M.}\ \bibnamefont {Rabe}},
  \bibinfo {author} {\bibfnamefont {E.}~\bibnamefont {Kaxiras}}, \ and\
  \bibinfo {author} {\bibfnamefont {J.~D.}\ \bibnamefont {Joannopoulos}},\
  }\href {\doibase 10.1103/PhysRevB.41.1227} {\bibfield  {journal} {\bibinfo
  {journal} {Phys. Rev. B}\ }\textbf {\bibinfo {volume} {41}},\ \bibinfo
  {pages} {1227} (\bibinfo {year} {1990})}\BibitemShut {NoStop}%
\bibitem [{\citenamefont {Ramer}\ and\ \citenamefont
  {Rappe}(1999)}]{Ramer99p12471}%
  \BibitemOpen
  \bibfield  {author} {\bibinfo {author} {\bibfnamefont {N.~J.}\ \bibnamefont
  {Ramer}}\ and\ \bibinfo {author} {\bibfnamefont {A.~M.}\ \bibnamefont
  {Rappe}},\ }\href {\doibase 10.1103/PhysRevB.59.12471} {\bibfield  {journal}
  {\bibinfo  {journal} {Phys. Rev. B}\ }\textbf {\bibinfo {volume} {59}},\
  \bibinfo {pages} {12471} (\bibinfo {year} {1999})}\BibitemShut {NoStop}%
\bibitem [{\citenamefont {Grinberg}\ \emph {et~al.}(2000)\citenamefont
  {Grinberg}, \citenamefont {Ramer},\ and\ \citenamefont
  {Rappe}}]{Ilya00p2311}%
  \BibitemOpen
  \bibfield  {author} {\bibinfo {author} {\bibfnamefont {I.}~\bibnamefont
  {Grinberg}}, \bibinfo {author} {\bibfnamefont {N.~J.}\ \bibnamefont {Ramer}},
  \ and\ \bibinfo {author} {\bibfnamefont {A.~M.}\ \bibnamefont {Rappe}},\
  }\href {\doibase 10.1103/PhysRevB.62.2311} {\bibfield  {journal} {\bibinfo
  {journal} {Phys. Rev. B}\ }\textbf {\bibinfo {volume} {62}},\ \bibinfo
  {pages} {2311} (\bibinfo {year} {2000})}\BibitemShut {NoStop}%
\bibitem [{Note2()}]{Note2}%
  \BibitemOpen
  \bibinfo {note} {See the supplementary material for the discussion on the
  doping concentration.}\BibitemShut {Stop}%
\bibitem [{\citenamefont {Jackiw}\ and\ \citenamefont
  {Rebbi}(1976)}]{Jackiw76p3398}%
  \BibitemOpen
  \bibfield  {author} {\bibinfo {author} {\bibfnamefont {R.}~\bibnamefont
  {Jackiw}}\ and\ \bibinfo {author} {\bibfnamefont {C.}~\bibnamefont {Rebbi}},\
  }\href {\doibase 10.1103/PhysRevD.13.3398} {\bibfield  {journal} {\bibinfo
  {journal} {Phys. Rev. D}\ }\textbf {\bibinfo {volume} {13}},\ \bibinfo
  {pages} {3398} (\bibinfo {year} {1976})}\BibitemShut {NoStop}%
\bibitem [{\citenamefont {Wang}\ \emph {et~al.}(2013)\citenamefont {Wang},
  \citenamefont {Weng}, \citenamefont {Wu}, \citenamefont {Dai},\ and\
  \citenamefont {Fang}}]{Wang13p125427}%
  \BibitemOpen
  \bibfield  {author} {\bibinfo {author} {\bibfnamefont {Z.}~\bibnamefont
  {Wang}}, \bibinfo {author} {\bibfnamefont {H.}~\bibnamefont {Weng}}, \bibinfo
  {author} {\bibfnamefont {Q.}~\bibnamefont {Wu}}, \bibinfo {author}
  {\bibfnamefont {X.}~\bibnamefont {Dai}}, \ and\ \bibinfo {author}
  {\bibfnamefont {Z.}~\bibnamefont {Fang}},\ }\href {\doibase
  10.1103/PhysRevB.88.125427} {\bibfield  {journal} {\bibinfo  {journal} {Phys.
  Rev. B}\ }\textbf {\bibinfo {volume} {88}},\ \bibinfo {pages} {125427}
  (\bibinfo {year} {2013})}\BibitemShut {NoStop}%
\bibitem [{\citenamefont {Xie}\ \emph {et~al.}(2015)\citenamefont {Xie},
  \citenamefont {Schoop}, \citenamefont {Seibel}, \citenamefont {Gibson},
  \citenamefont {Xie},\ and\ \citenamefont {Cava}}]{Xie15p1}%
  \BibitemOpen
  \bibfield  {author} {\bibinfo {author} {\bibfnamefont {L.~S.}\ \bibnamefont
  {Xie}}, \bibinfo {author} {\bibfnamefont {L.~M.}\ \bibnamefont {Schoop}},
  \bibinfo {author} {\bibfnamefont {E.~M.}\ \bibnamefont {Seibel}}, \bibinfo
  {author} {\bibfnamefont {Q.~D.}\ \bibnamefont {Gibson}}, \bibinfo {author}
  {\bibfnamefont {W.}~\bibnamefont {Xie}}, \ and\ \bibinfo {author}
  {\bibfnamefont {R.~J.}\ \bibnamefont {Cava}},\ }\href@noop {} {\enquote
  {\bibinfo {title} {{Potential ring of Dirac nodes in a new polymorph of
  Ca$_3$P$_2$}},}\ } (\bibinfo {year} {2015}),\ \Eprint
  {http://arxiv.org/abs/arXiv:1504.01731} {arXiv:1504.01731} \BibitemShut
  {NoStop}%
\bibitem [{\citenamefont {Wang}\ \emph {et~al.}(2012)\citenamefont {Wang},
  \citenamefont {Sun}, \citenamefont {Chen}, \citenamefont {Franchini},
  \citenamefont {Xu}, \citenamefont {Weng}, \citenamefont {Dai},\ and\
  \citenamefont {Fang}}]{Wang12p195320}%
  \BibitemOpen
  \bibfield  {author} {\bibinfo {author} {\bibfnamefont {Z.}~\bibnamefont
  {Wang}}, \bibinfo {author} {\bibfnamefont {Y.}~\bibnamefont {Sun}}, \bibinfo
  {author} {\bibfnamefont {X.-Q.}\ \bibnamefont {Chen}}, \bibinfo {author}
  {\bibfnamefont {C.}~\bibnamefont {Franchini}}, \bibinfo {author}
  {\bibfnamefont {G.}~\bibnamefont {Xu}}, \bibinfo {author} {\bibfnamefont
  {H.}~\bibnamefont {Weng}}, \bibinfo {author} {\bibfnamefont {X.}~\bibnamefont
  {Dai}}, \ and\ \bibinfo {author} {\bibfnamefont {Z.}~\bibnamefont {Fang}},\
  }\href {\doibase 10.1103/PhysRevB.85.195320} {\bibfield  {journal} {\bibinfo
  {journal} {Phys. Rev. B}\ }\textbf {\bibinfo {volume} {85}},\ \bibinfo
  {pages} {195320} (\bibinfo {year} {2012})}\BibitemShut {NoStop}%
\bibitem [{\citenamefont {Teo}\ and\ \citenamefont
  {Kane}(2010)}]{Teo10p115120}%
  \BibitemOpen
  \bibfield  {author} {\bibinfo {author} {\bibfnamefont {J.~C.~Y.}\
  \bibnamefont {Teo}}\ and\ \bibinfo {author} {\bibfnamefont {C.~L.}\
  \bibnamefont {Kane}},\ }\href {\doibase 10.1103/PhysRevB.82.115120}
  {\bibfield  {journal} {\bibinfo  {journal} {Phys. Rev. B}\ }\textbf {\bibinfo
  {volume} {82}},\ \bibinfo {pages} {115120} (\bibinfo {year}
  {2010})}\BibitemShut {NoStop}%
\end{thebibliography}%

\end{document}